\documentclass[12pt]{article}
\usepackage{epsf}
\usepackage{amsmath,amssymb}

\usepackage{graphicx}
\usepackage{comment}
\usepackage{mathrsfs}

\usepackage{tikz}
\usetikzlibrary{decorations.pathmorphing}
\usetikzlibrary{decorations.markings}
\usepackage{bm}
\usepackage{cite}
\usepackage{hyperref}

\begin{document}

\newcommand{\lsim}{\stackrel{<}{_\sim}}
\newcommand{\gsim}{\stackrel{>}{_\sim}}
\newcommand{\mathhyphen}{\mathchar"712D}

\renewcommand{\theequation}{\thesection.\arabic{equation}}

\renewcommand{\thefootnote}{\fnsymbol{footnote}}
\setcounter{footnote}{0}

\begin{titlepage}

\begin{center}

\hfill February, 2021

\vskip .5in

{\Large\bf
  Leptonic CP and Flavor Violations
  \\[2mm]
  in SUSY GUT with Right-handed Neutrinos
  }

\vskip .5in

{\large
  Kaigo Hirao and Takeo Moroi
}

\vskip 0.25in
{\em Department of Physics, University of Tokyo, Tokyo 113-0033, Japan}

\end{center}

\vskip .3in

\begin{abstract}

  We study leptonic CP and flavor violations in supersymmetric (SUSY)
  grand unified theory (GUT) with right handed neutrinos,
  paying attention to the renormalization group effects on
  the slepton mass matrices due to the neutrino and GUT Yukawa
  interactions. In particular, we study in detail the impacts of the so-called
  Casas-Ibarra parameters on CP and flavor violating observables.
  The renormalization group effects induce CP and flavor
  violating elements of the SUSY breaking scalar mass squared
  matrices, which may result in sizable leptonic CP and flavor
  violating signals.  Assuming seesaw formula for the active neutrino
  masses, the renormalization group effects have been often thought to
  be negligible as the right-handed neutrino masses become small.
  With the most general form of the neutrino Yukawa matrix, i.e.,
  taking into account the Casas-Ibarra parameters, however,
  this is not the case.  We found that the maximal possible sizes of
  signals of leptonic CP and flavor violating processes are found to
  be insensitive to the mass scale of the right-handed neutrinos and
  that they are as large as (or larger than) the present experimental
  bounds irrespective of the right-handed neutrino masses.

\end{abstract}

\end{titlepage}

\setcounter{page}{1}
\renewcommand{\thefootnote}{\#\arabic{footnote}}
\setcounter{footnote}{0}

\section{Introduction}
\setcounter{equation}{0}

Even though the standard model (SM) of particle physics successfully
explains many of results of high energy experiments, the existence of
a physics beyond the SM (BSM) has been highly anticipated.
Particularly, from particle cosmology point of view, there are many
miseries which cannot be explained in the framework of the SM, like
the existence of dark matter, the origin of the baryon asymmetry of
the universe, the dynamics of inflation, and so on.  Many experimental
efforts have been performed to find signals of the BSM physics.

In the search of the BSM signals, energy and precision frontier
experiments are both important.  The energy frontier experiments,
represented by collider experiments like the LHC at present, may
directly find and study particles in BSM models, but their discovery
reach is limited by the beam energy.  On the contrary, the precision
frontier ones may reach the BSM whose energy scale is much higher than
the energy scale of the LHC experiment, although information about the
BSM from those experiments may be indirect.  Currently, the LHC has
not found any convincing evidence of the BSM physics.  In such a
circumstance, it is important to reconsider the role of precision
frontier experiments and study what kind of signal may be obtained
from them.

In this paper, we study CP and flavor violations in models with
supersymmetry (SUSY), and their impacts on on-going and future
experiments.  Even though the LHC has not found any signal of SUSY
particles with their mass scale of $\sim$ TeV, the SUSY is still a
well-motivated candidate of BSM physics.  Taking into account the
observed Higgs mass of $125.10\ {\rm GeV}$ \cite{Zyla:2020zbs}, heavy
SUSY particles (more specifically, heavy stops) are preferred to push
up the Higgs mass via radiative corrections \cite{Okada:1990vk,
  Okada:1990gg, Ellis:1990nz,Haber:1990aw}.  We note that, in a large
class of models, SUSY particles can acquire masses of $\sim
O(10-100)\, {\rm TeV}$ \cite{Wells:2004di, Giudice:1998xp, Ibe:2006de,
  Ibe:2011aa, ArkaniHamed:2012gw}.  Here, we pay particular attention
to SUSY SU(5) grand unified theory (GUT) with right handed neutrinos,
in which superparticle masses are much above the TeV scale, because
(i) with the particle content of the minimal SUSY standard model, the
gauge coupling unification at $M_{\rm GUT}\sim O(10^{16})\ {\rm GeV}$
is suggested, and also because (ii) right-handed neutrinos are
well-motivated to explain the origin of the active neutrino masses via
the seesaw mechanism \cite{Minkowski:1977sc,
  Yanagida:1979as, GellMann:1980vs}.  In SUSY GUT with right handed
neutrinos, some of the cosmological mysteries mentioned above may be
also solved; the baryon asymmetry of the universe may be explained by
the leptogenesis scenario \cite{Fukugita:1986hr}, while the lightest
superparticle (LSP) may play the role of dark matter.  Compared to the
SM, the SUSY models contain various new sources of CP and flavor
violations.  It may cause significant CP and flavor violating
processes which cannot be explained in the SM.  If such processes are
experimentally observed, they can be smoking gun evidences of the BSM
physics, based on which we may study the BSM model behind the CP and
flavor violations.

It has been well known that the renormalization group effect may
induce CP and flavor violating off diagonal elements of the slepton
mass matrices \cite{Borzumati:1986qx, Barbieri:1994pv,
  Barbieri:1995tw, Romanino:1996cn}; in the framework of our interest,
the left and right handed slepton mass matrices are affected by the
renormalization group effects from the neutrino Yukawa coupling and
the running above the GUT scale $M_{\rm GUT}$, respectively, even
though the lepton flavor is conserved in the Yukawa interaction of the
minimal SUSY standard model (MSSM).  Thus, even though the slepton
mass matrices are universal at some high scale (for example, Planck
scale), such universalities are violated by the renormalization group
effects.  The effects of the off diagonal elements of the slepton mass
matrices on CP and/or flavor violating observables have been studied
(see, for example, \cite{Hisano:1995nq, Hisano:1995cp, Hisano:1998fj,
  Baek:2000sj, Moroi:2000mr, Moroi:2000tk, Casas:2001sr, Akama:2001em,
  Ellis:2001xt, Ellis:2001yza, Chang:2002mq, Ellis:2002fe, Hisano:2003bd,
  Masina:2003wt, Ciuchini:2003rg, Hisano:2004pw, Calibbi:2006nq,
  Hisano:2008df, Hisano:2008hn, Borzumati:2009hu, Moroi:2013sfa,
  McKeen:2013dma, Moroi:2013vya, Altmannshofer:2013lfa, Smith:2017dtz,
  Evans:2018ewb}).  In particular, it has been pointed out that, even
if the MSSM particles are out of the reach of the LHC experiment, the
signal of the MSSM may be observed by on-going or future CP or flavor
violation experiments.  In previous studies, the neutrino Yukawa
matrix was reconstructed by combining the seesaw formula with the
active neutrino mass squared differences suggested by the neutrino
oscillation experiments.  Then, the neutrino Yukawa coupling constants
are inversely proportional to the square root of the mass scale of the
right handed neutrinos.  With adopting simple assumption about the
neutrino sector, i.e., the universal masses for the right handed
neutrinos as well as a simple mixing structure, the renormalization
group effects due to the neutrino Yukawa couplings become irrelevant
as the mass scale of the right handed neutrinos becomes smaller.
However, as pointed out by Casas and Ibarra (CI) \cite{Casas:2001sr},
there exist several parameters (which we call CI parameters) which
complicate the mixing structure of the neutrino Yukawa matrix.

In this paper, we study CP and flavor violating processes paying
particular attention to the effects of the CI parameters, as well as
the effects of the non-universality of the right handed neutrino
masses, whose effects have not been fully investigated so far.  (For
some discussion about the effect of the CI parameters, see
\cite{Hisano:1998fj, Ellis:2001xt, Ellis:2001yza, Ellis:2002fe,
Hisano:2003bd, Calibbi:2006nq, Smith:2017dtz}). 
The organization of this paper is as follows.
In Section \ref{sec:model}, we introduce the model based on which we
perform our analysis. In Section \ref{sec:numerical}, we show the
results of our numerical analysis.  Section \ref{sec:conclusion} is
devoted to conclusions and discussion.

\section{Model and Parameterization}
\label{sec:model}
\setcounter{equation}{0}

In this section, we introduce the model we consider.  We also
summarize our convention of the model parameters, including
CI parameters and GUT phases.  To this end, we define the
couplings and specify the flavor basis we use for each effective
theories and explain how they are related at the matching scales.

\begin{figure}[t]
\centering
\begin{tikzpicture}[baseline= {([yshift=-1ex]current bounding box.center)}]
\coordinate(A)at(0,0);\coordinate(G)at(-2.0,0);
\draw(G)--node[midway,above]{QEDQCD}(A)
--++(0.5,0)coordinate(B)node[below]{$M_{t}$}
--++(2.5,0)coordinate(C)node[below]{$M_{S}$}node[midway,above]{SM}
--++(3.5,0)coordinate(D)node[below]{$$}node[midway,above]{MSSM}
--++(0.25,0)coordinate(H)node[below]{$\underline{M}_{N_{1,2,3}}$}
--++(0.25,0)coordinate(I)node[below]{$$}
--++(3,0)coordinate(E)node[below]{$M_{\rm{GUT}}$}node[midway,above]{MSSMNR}
--++(3,0)coordinate(F)node[below]{$M_{\rm{Pl}}$}node[midway,above]{SU(5)NR};
\draw[->](F)--++(0.5,0)node[right,scale=1.0]{$Q$}node[below right,scale=0.8]{$$};
\draw(B)--++(0,0.1);
\draw(B)--++(0,-0.1);
\draw(C)--++(0,0.1);
\draw(C)--++(0,-0.1);
\draw(D)--++(0,0.1);
\draw(D)--++(0,-0.1);
\draw(E)--++(0,0.1);
\draw(E)--++(0,-0.1);
\draw(F)--++(0,0.1);
\draw(F)--++(0,-0.1);
\draw(H)--++(0,0.1);
\draw(H)--++(0,-0.1);
\draw(I)--++(0,0.1);
\draw(I)--++(0,-0.1);
\end{tikzpicture}
\caption{The effective theories used for each regions of the
  renormalization scale $Q$.}
\label{fig:eft}
\end{figure}
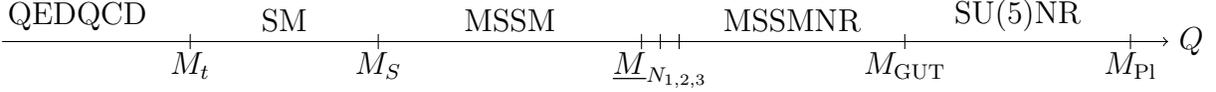

Effective theories in our model appropriate for the each energy scales
lower than $M_{\rm Pl}$ are shown in Fig.\ \ref{fig:eft}, where
\begin{itemize}
\item QEDQCD: QED and QCD
\item MSSMNR: MSSM with three generations of right-handed neutrinos
\item SU(5)NR: minimal SU(5) GUT with three generations of
  right-handed neutrinos
\end{itemize}
At each renormalization scale $Q$, we use the relevant effective theory
as we explain below.

We assume that the effect of the SUSY breaking is mediated to the
visible sector (containing the MSSM particles and right-handed
neutrinos) at the reduced Planck scale $M_{\rm Pl}\simeq 2.4\times
10^{18}\, {\rm GeV}$.  Then, at the scales between $M_{\rm Pl}$ and
$M_{\rm GUT}$, the model is described by SU(5)NR.  In order to
introduce three generations of quarks and leptons, three copies of
chiral supermultiplets $\Phi_i$ and $\Psi_i$, which are in the
$\bm{\bar{5}}$ and $\bm{10}$ representations of SU(5),
respectively, are introduced.  (Here, the $i=1-3$ is the generation
index.)  As the conventional SU(5) GUT, $\Phi_i$ is composed of the
right-handed down-type quark multiplets $\bar{D}_i$ and the lepton
doublets $L_i$, while $\Psi_i$ is composed of the quark doublets
$Q_i$, the right handed up-type quark multiplets $U_i$ and the
right-handed charged lepton $\bar{E}_i$.  The right-handed neutrinos
$\bar{N}_i$ of MSSMNR are added as SU(5) singlets $\Upsilon_i$. The
MSSM Higgs doublets $H_u$ and $H_d$ are contained embedded into $H$
and $\bar{H}$, which are SU(5) $\bm{5}$ and $\bm{\bar{5}}$
representations, respectively. There is also a multiplet which breaks
SU(5) symmetry to the SM gauge group.  We assume that a chiral
multiplet in the adjoint representation of SU(5), which we call
$\Sigma$, is responsible for the breaking of the SU(5) symmetry.  The
vacuum expectation value (VEV) of $\Sigma$ is denoted as
$\langle\Sigma\rangle={\rm diag}(2v_{\rm GUT},2v_{\rm GUT},2v_{\rm
  GUT},-3v_{\rm GUT},-3v_{\rm GUT})$.

We consider the superpotential of SU(5)NR in the following form:
\begin{align}
  W_{\rm SU(5)NR}&=W_{\rm SU(5)NR}^{\rm ren}+W_{\rm SU(5)NR}^{\rm nonren},
  \label{Wsu5nr}\\
  W_{\rm SU(5)NR}^{\rm ren}&=W_{\rm SU(5)NR}^{\rm matter}+W_{\rm SU(5)NR}^{\rm Higgs},
  \label{Wsu5nrren}
\end{align}
and 
\begin{align}  
  W_{\rm SU(5)NR}^{\rm matter}&=\frac{1}{4}(f_u)_{ij}\Psi_{i}\Psi_{j}H
  +\sqrt{2}(f_d)_{ij}\Psi_{i}\Phi_{j}\bar{H}
  +(f_{\nu})_{ij}\Upsilon_{i}\Phi_{j}H
  +\frac{1}{2}(M_{\Upsilon})_{ij}\Upsilon_i\Upsilon_j,
\label{Wsu5nrrenmat}
\end{align}
where $f_u$, $f_d$, and $f_{\nu}$ are $3\times 3$ coupling matrices
while $M_{\Upsilon}$ is $3\times 3$ matrix with mass dimension $1$.
Notice that $f_u$ and $M_{\Upsilon}$ are symmetric.  In
Eq.\ \eqref{Wsu5nrrenmat}, the summations over SU(5) indices are
implicit.  (We follow \cite{Borzumati:2009hu} for the group theoretical
notations.)  $W_{\rm SU(5)NR}$ consists of the renormalizable part
$W_{\rm SU(5)NR}^{\rm ren}$ and the non-renormalizable part $W_{\rm
  SU(5)NR}^{\rm nonren}$. $W_{\rm SU(5)NR}^{\rm ren}$ is further
split into $W_{\rm SU(5)NR}^{\rm matter}$ (i.e., the superpotential
containing the matter sector) and $W_{\rm SU(5)NR}^{\rm Higgs}$ (i.e.,
the superpotential for the Higgs sector); $W_{\rm SU(5)NR}^{\rm
  Higgs}$ is the superpotential containing only the Higgs field and
$\Sigma$.  In Eq.\ \eqref{Wsu5nrrenmat}, $W_{\rm SU(5)NR}^{\rm matter}$
contains superpotential responsible for the up-type, down-type and
neutrino-type Yukawa terms in the MSSMNR.  In addition, in order to
explain the unification of the down-type and electron-type Yukawa
matrices, we assume that $W_{\rm SU(5)NR}^{\rm nonren}$ contains a
term in the following form:
\begin{align}  
  W_{\rm SU(5)NR}^{\rm nonren} \ni
  \frac{\sqrt{2}}{M_{\rm Pl}} c_{ij} \Psi_{i} \Sigma \Phi_{j} \bar{H}.
\end{align}

Unitary rotations on the family indices can make the coupling matrices
to the following forms:
\begin{align}
  f_u&=V^T\hat{f}_u\hat{\Theta}_qV,\\
  \label{f_u}
  f_d&=\hat{f}_d,\\
  f_{\nu}&= \tilde{W}^{\dagger}\hat{f}_{\nu}U^{\dagger}\hat{\Theta}_l,
  \label{highpar}\\
  M_{\Upsilon}&=\hat{M}_{\Upsilon},
  \label{Mupsilon}
\end{align}
where $\hat{f}_{u}$, $\hat{f}_{d}$, $\hat{f}_{\nu}$ and $\hat{M}_{\Upsilon}$ are
real diagonal matrices.\footnote
{The hat on matrix symbols indicates that they are diagonal.}
In addition, $V$ and $U$ are unitary matrices with only a single CP
phase and three mixing angles while $\tilde{W}$ is a general unitary
matrix with additional 5 phases.\footnote
{In general, a unitary matrix $\tilde{X}$ can be
  decomposed as
  \begin{align*}
    \tilde{X} =
    e^{i\varphi_{\tilde{X}}}
    \hat{\Theta}^{(L)}_{{\tilde{X}}}
    X
    \hat{\Theta}^{(R)}_{{\tilde{X}}},
  \end{align*}
  where $\varphi_{\tilde{X}}$ is the overall phase of the
  matrix ${\tilde{X}}$,
  $\hat{\Theta}^{(L)}_{{\tilde{X}}}$ and
  $\hat{\Theta}^{(R)}_{{\tilde{X}}}$ are diagonal phase
  matrices parameterized by two physical phases, and $X$ is
  a unitary matrix parameterized by three mixing angles,
  $\vartheta_{12}$, $\vartheta_{13}$, and $\vartheta_{23}$, and a
  single phase $\delta$ as
  \begin{align*}
    X =
    \left( \begin{array}{ccc}
      c_{12} c_{13} & s_{12} c_{13} & s_{13} e^{-i\delta} \\
      -s_{12}c_{23}-c_{12}s_{23}s_{13} e^{i\delta} & 
      c_{12}c_{23}-s_{12}s_{23}s_{13} e^{i\delta} & 
      s_{23}c_{13}\\
      s_{12}s_{23}-c_{12}c_{23}s_{13} e^{i\delta} & 
      -c_{12}c_{23}-s_{12}c_{23}s_{13} e^{i\delta} & 
      c_{23}c_{13}
    \end{array} \right),
  \end{align*}
  with $c_{ij}=\cos\vartheta_{ij}$ and $s_{ij}=\sin\vartheta_{ij}$.}
Furthermore, $\hat{\Theta}_{q}$ and $\hat{\Theta}_{l}$ are diagonal
phase matrices and represent CP phases intrinsic in SU(5) GUT.  Notice
that the overall phases of $\hat{\Theta}_{q}$ and $\hat{\Theta}_{l}$
are unphysical because they can be absorbed to $\tilde{W}$.  Thus,
each of $\hat{\Theta}_{q}$ and $\hat{\Theta}_{l}$ contains two
parameters; we parameterize these matrices as
\begin{align}
  \hat{\Theta}_{f}= {\rm{diag}}
  (1,  e^{i\varphi_{f_2}}, e^{i\varphi_{f_3}}),
\end{align}
with $f=q,l$.  

In the following argument, we take the flavor basis in which the
coupling matrices of SU(5)NR take the forms of Eqs.\ \eqref{f_u} --
\eqref{Mupsilon} at $Q=M_{\rm GUT}$.  In our discussion, the higher
dimensional operator proportional to $c$ is introduced just to
guarantee the unification of $\bar{U}_i$ and $L_i$ into $\bm{\bar{5}}$
multiplets of SU(5).  For simplicity, we assume that $c$ is real and
diagonal at $Q=M_{\rm GUT}$ in this basis: $c=\hat{c}$.

At the GUT scale $M_{\rm GUT}$, SU(5)NR couplings are matched to
MSSMNR couplings.  The matter sector superpotential of MSSMNR is given
in the following form:
\begin{align}
  W_{{\rm{MSSMNR}}}^{\rm matter}=
  (y_u)_{ij}H_u\bar{U}_i Q_j +(y_d)_{ij}H_d\bar{D}_i Q_j
  + (y_{e})_{ij}H_d \bar{E}_iL_j +(y_{\nu })_{ij}H_u\bar{N}_iL_j
  + \frac{1}{2}(M_N)_{ij}\bar{N}_i\bar{N}_j,
\label{Wmssmnr}
\end{align}
where $y_u$, $y_d$, $y_e$ and $y_{\nu}$ are the MSSMNR Yukawa matrices
while $M_N$ is the Majorana mass matrix of $\bar{N}$.  The MSSMNR
chiral multiplets are embedded into the SU(5)NR ones as follows:
\begin{align}
  \Psi_i &=
  (\mathcal{U}_Q^\dagger Q,
  V^{\dagger}\hat{\Theta}_{q}^{*} \mathcal{U}_{\bar U}^\dagger \bar{U},
  \hat{\Theta}_l \mathcal{U}_{\bar E}^\dagger \bar{E})_i,
  \\
  \Phi_i &= (\mathcal{U}_{\bar D}^\dagger \bar{D},
  \hat{\Theta}_l^* \mathcal{U}_L^\dagger  L)_i,
  \\
  \Upsilon_i &=(\mathcal{U}_{\bar N}^\dagger \bar{N})_i,
\end{align}
where $\mathcal{U}_Q$, $\mathcal{U}_{\bar U}$, $\mathcal{U}_{\bar
  D}$, $\mathcal{U}_L$, $\mathcal{U}_{\bar E}$, and
$\mathcal{U}_{\bar N}$ are $3\times 3$ unitary matrices which depend
on the choice of the flavor basis in the MSSMNR.  At the tree level,
the matching conditions at $Q=M_{\rm GUT}$ are obtained as
\begin{align}
  \mathcal{U}_{\bar U}^T \left[ y_u \right]_{Q=M_{\rm GUT}} \mathcal{U}_Q &=
  \left[ \hat{f}_u V \right]_{Q=M_{\rm GUT}},
  \label{yuMgut}\\
  \mathcal{U}_{\bar D}^T \left[ y_d \right]_{Q=M_{\rm GUT}} \mathcal{U}_Q &=
  \left[ \hat{f}_d + \frac{2v_{\rm GUT}}{M_{\rm Pl}}\hat{c} \right]_{Q=M_{\rm GUT}},
  \label{ydMgut}\\
  \mathcal{U}_{\bar E}^T \left[ y_e \right]_{Q=M_{\rm GUT}} \mathcal{U}_L &= 
  \left[ \hat{f}_{d} -\frac{3v_{\rm GUT}}{M_{\rm Pl}}\hat{c} \right]_{Q=M_{\rm GUT}},
  \label{yeMgut}\\
  \mathcal{U}_{\bar N}^T \left[ y_{\nu} \right]_{Q=M_{\rm GUT}} \mathcal{U}_L &=
  \left[ \tilde{W}^\dagger \hat{f}_{\nu} U^\dagger \right]_{Q=M_{\rm GUT}},
  \label{ynMgut}\\
  \mathcal{U}_{\bar N}^T \left[ M_{N} \right]_{Q=M_{\rm GUT}}\mathcal{U}_{\bar N}
  &=
  \left[ \hat{M}_{\Upsilon} \right]_{Q=M_{\rm GUT}}.
  \label{MnMgut}
\end{align}
Based on the above relations, the coupling matrices of the SU(5)NR are
determined from those of the MSSMNR in our numerical analysis with
properly choosing the unitary matrices $\mathcal{U}_Q$,
$\mathcal{U}_{\bar U}$, $\mathcal{U}_{\bar D}$, $\mathcal{U}_L$,
$\mathcal{U}_{\bar E}$, and $\mathcal{U}_{\bar N}$.  Notice that the
GUT phases $\hat{\Theta}_q$ and $\hat{\Theta}_l$ can be absorbed into
the definitions of the MSSMNR superfields and can be removed from the
MSSMNR superpotential; however, they are physical in SU(5)NR.

Now, let us consider the neutrino masses.  For this purpose, it is
more convenient to use the flavor basis in which $\bar{N}_i$ ($i=1-3$)
become the mass eigenstates (see below).  In general, the masses of
right-handed neutrinos are different.  We denote the mass of $i$-th
right-handed neutrino as $\underline{M}_{N_i}$, with
$\underline{M}_{N_1}\leq\underline{M}_{N_2}\leq\underline{M}_{N_3}$.
The dimension-five operator responsible for the Majorana mass terms of
the left-handed neutrinos is generated with see-saw mechanism by
integrating out right-handed neutrinos \cite{Yanagida:1979as, GellMann:1980vs,
Minkowski:1977sc}.  Let us define the
following diagonal matrix: $\hat{\underline{M}}_N\equiv{\rm
  diag}(\underline{M}_{N_1},
\underline{M}_{N_2},\underline{M}_{N_3})$.  Using
$\hat{\underline{M}}_N$, the active neutrino mass matrix is given
by\footnote
{In Eq.\ \eqref{seesawformula}, the running of the Wilson coefficients of
  the dimension-five operator is neglected in our analysis. }
\begin{align}
  m_{\nu} = v_u^2
  \underline{y}_{\nu}^{T}
  \hat{\underline{M}}_N^{-1} \underline{y}_{\nu}
\label{seesawformula}
\end{align}
where $v_u$ is the VEV of $H_u$.  In addition,
$\underline{y}_{\nu,ij}\equiv y_{\nu,ij}(Q=\underline{M}_{N_i})$,
with $j=1-3$,  is the neutrino Yukawa coupling constant at
$Q=\underline{M}_{N_i}$, where, again, we are adopting the flavor
basis in which $N_i$ is the mass eigenstate.

At the energy scales lower than $\underline{M}_{N_1}$, the model is
described by the MSSM and one can always work in the basis in which
$y_e$ is diagonal.  In such a basis, $m_{\nu}$ takes the
following form.
\begin{align}
  m_{\nu} &=U_{\rm PMNS}^* \hat{m}_{\nu}\hat{\Theta}_M^2 U_{\rm PMNS}^{\dagger},
  \label{seesawconstraint}
\end{align}
where $U_{\rm PMNS}$ is the Pontecorvo-Maki-Nakagawa-Sakata (PMNS)
matrix \cite{Maki:1962mu, Pontecorvo:1967fh} with three mixing angles
$\theta_{12}$, $\theta_{23}$, $\theta_{13}$, and a Dirac CP phase
$\delta_{\rm{CP}}$, while diagonal phase matrix $\hat{\Theta}_M$
contains Majorana phases.  The overall phase of $\hat{\Theta}_M$ is
unphysical, and we adopt the convention such that $\hat{\Theta}_M={\rm
  diag}(1,e^{i\varphi_{M_2}},e^{i\varphi_{M_3}})$.  Furthermore,
$\hat{m}_{\nu}$ is a real diagonal matrix containing mass eigenvalues
of left-handed neutrinos.  In Table \ref{tab:pmns}, we summarize the
parameters in $\hat{m}_{\nu}$ and $U_{\rm PMNS}$ used in our numerical
analysis.  Here, we assume the normal hierarchy for neutrino masses
and take $m^2_{\nu_1}=1.00\times 10^{-6}\, {\rm eV}^2$.

\begin{table}[t]
\centering
\begin{tabular}{|c|c|c|c|c|c|c|}\hline
$\sin^2{\theta_{12}}$&$\sin^2{\theta_{23}}$&$\sin^2{\theta_{13}}$&$\delta_{\rm CP}$&
$\Delta m^2_{31}\, \left[\mathrm{eV^2}\right]$&$\Delta m^2_{21}\, \left[\mathrm{eV^2}\right]$&
$m_{\nu_1}^2\, \left[\mathrm{eV^2}\right]$\\\hline
$0.307$&$0.545$&$0.0218$&$1.36\pi$&$2.453\times 10^{-3}$&$7.53\times 10^{-5}$&$1.00\times 10^{-6}$\\\hline
\end{tabular}
\caption{Model parameters used in our numerical analysis \cite{Zyla:2020zbs}.}
\label{tab:pmns}
\end{table}

Comparing Eq.\ \eqref{seesawformula} and
Eq.\ \eqref{seesawconstraint}, $\underline{y}_\nu$
can be expressed in the following form, i.e., the so-called
Casas-Ibarra parameterization \cite{Casas:2001sr}:\footnote
{Square root of a diagonal matrix is understood to be applied to each diagonal
  elements.}
\begin{align}
  \underline{y}_{\nu}=\frac{1}{v_u}
  \hat{\underline{M}}_N^{\frac{1}{2}}R
  \hat{m}_{\nu}^{\frac{1}{2}}\hat{\Theta}_MU_{\rm{PMNS}}^{\dagger},
  \label{lowpar}
\end{align}
where $R$ is an arbitrary complex and orthogonal matrix:
\begin{align}
  RR^T=\bm{1}.
\end{align}
The two parameterizations of the neutrino Yukawa matrices,
Eq.\ \eqref{ynMgut} and Eq.\ \eqref{lowpar}, are equivalent. The
number of parameters in both parameterizations are summarized in Tables
\ref{tab:paramhigh} and \ref{tab:paramCI}.  (Notice that the overall
phase of the unitary matrix $\tilde{W}$, as well as the phase matrices
$\hat{\Theta}_q$ and $\hat{\Theta}_l$, cannot be determined from the
low energy observables.  We call them ``GUT phases.'')

The complex orthogonal matrix $R$ can be decomposed into the product
of a real orthogonal matrix $O$ and a hermitian and orthogonal matrix
$H$ as $R=OH$.  Let us define
\begin{align}
  n(\theta,\phi)\equiv(\sin{\theta}\cos{\phi},\sin{\theta}\sin{\phi},\cos{\theta}),
\end{align}
and 
\begin{align}
  (A(n))_{ij}&\equiv \epsilon_{ijk}n_k.
\end{align}
Then, $H$ can be expressed as
\begin{align}
  H(r,n) = e^{irA(n)} = 
  nn^{T} + \cosh{r}\left(\bm{1}-nn^{T} \right) + i\sinh{r} A(n),
\end{align}
and hence is parameterized by 3 real parameters $(r,\theta,\phi)$.  We
can derive another useful expression of $H(r,n)$.  For this purpose,
we introduce a complex vector
\begin{align}
\tilde{n}\equiv \frac{1}{\sqrt{2}}(n'-in''),
\end{align}
where $n'$ and $n''$ are an arbitrary set of 2 unit vectors such that 
$\langle n,n',n''\rangle$ forms a right-handed orthonormal basis of 
$\mathbb{R}^3$. One can easily check that 
$\langle n,\tilde{n},\tilde{n}^*\rangle$ forms an orthonormal basis of
$\mathbb{C}^3$. With these vectors, $H(r,n)$ can be expressed as
\begin{align}
  H(r,n) &=
  e^r \tilde{P}(n)+P(n)+e^{-r}\tilde{P}^{*}(n),
  \label{Hmatrix}
\end{align}
where
\begin{align}
 P(n)\equiv nn^T,\ \tilde{P}(n)\equiv \tilde{n}\tilde{n}^{\dagger},
  \ \tilde{P}^*(n)\equiv \tilde{n}^*\tilde{n}^{T}.
\end{align}
Here $P$, $\tilde{P}$, and $\tilde{P}^*$ are orthogonal projection matrices
onto $\mathbb{C}n$, $\mathbb{C}\tilde{n}$, and
$\mathbb{C}\tilde{n}^*$, respectively. Note that $\tilde{P}$ depends 
only on $n$ and not on the choice of $n'$ and $n''$. When $r\gg 1$,
the first term in the right-hand side of Eq.\ \eqref{Hmatrix} dominates.

\begin{table}[t]
\centering
\begin{tabular}{|c|c|c|c|c|}\hline
$\hat{\Theta}_{l}$&$\tilde{W}$&$U$&$\hat{f}_{\nu}$&total\\\hline
$2$&$9$&$4$&$3$&18\\\hline
\end{tabular}
\caption{The number of real degrees of freedom in the neutrino Yukawa
  matrix in the parameterization of Eq.\ \eqref{ynMgut}.}
\label{tab:paramhigh}
\vspace{3mm}
\begin{tabular}{|c|c|c|c|c|c|c|c|}\hline
  $\varphi_{\tilde{W}}$&
  $\hat{\Theta}_{l}$&$\hat{\Theta}_{M}$&$U_{\rm{PMNS}}$&$\hat{m}_{\nu}$&$O$&$H$&total\\\hline
  $1$&
  $2$&$2$&$4$&$3$&$3$&$3$&18\\\hline
\end{tabular}
\caption{The number of real degrees of freedom in the neutrino Yukawa
  matrix in CI parameterization, where $\varphi_{\tilde{W}}$ is the
  overall phase of the matrix $\tilde{W}$. }
\label{tab:paramCI}
\end{table}

Many of the previous analysis of the flavor violations have not paid
significant attention to the effect of the CI parameters, taking
$R=\bm{1}$ (see, however, \cite{Hisano:1998fj, Ellis:2001xt,
Ellis:2001yza, Ellis:2002fe, Hisano:2003bd, Calibbi:2006nq,
Smith:2017dtz}). In addition, it has
been often assumed that the right-handed neutrino masses are
degenerate, i.e.,
$\underline{M}_{N_1}=\underline{M}_{N_2}=\underline{M}_{N_3}$.  With
these simplifications,
$\underline{y}_{\nu}=\frac{1}{v_u}(\hat{\underline{M}}_{N}\hat{m}_{\nu})^{\frac{1}{2}}U_{\rm
  PMNS}^{\dagger}$.  However, as we will see in the following,
parameters in $R$ may significantly affect the CP and flavor violating
observables.

\section{Numerical Analysis}
\label{sec:numerical}
\setcounter{equation}{0}

Now, let us numerically evaluate the CP and flavor violating
observables.  Our primary purpose is to study the effects of CI
parameters on electron electric dipole moment (EDM) and branching ratios of lepton flavor
violating (LFV) processes.  Thus, for simplicity, we assume that the
soft SUSY breaking parameters satisfy the so-called mSUGRA boundary
conditions; the soft scalar mass-squared parameters at the Planck
scale $Q=M_{\rm Pl}$ are assumed to be universal (and are equal to
$m_0^2$), and tri-linear scalar couplings (so-called $A$-terms) are
proportional to corresponding Yukawa couplings (with the
proportionality constant of $a_0$).

In our analysis, we calculate the MSSM parameters at the mass scale of
MSSM superparticles (which we call MSSM scale).  Here are remarks
about our calculation:
\begin{itemize}
\item The input SM parameters related to low energy observables are
  \begin{align}
    g_a,\ \hat{y}_u,\ \hat{y}_d,\ V_{\rm CKM},\ \hat{y}_e,\
    \hat{m}_{\nu},\ U_{\rm PMNS},
  \end{align}
  where $g_a$ (with $a=1-3$) are gauge coupling constants, while
  $V_{\rm CKM}$ is the Cabibbo-Kobayashi-Maskawa (CKM) matrix.  For
  the boundary conditions of SM couplings at the top mass scale
  $M_{t}$, we follow \cite{Buttazzo:2013uya}. The parameters in the
  CKM matrix are taken from \cite{Zyla:2020zbs} and set at $M_t$; the
  lightest neutrino mass, which cannot be determined from current
  neutrino oscillation experiments, is set to be $1\times 10^{-3}\,
  {\rm eV}$.  For the left-handed neutrino mass eigenstates and the
  PMNS matrix, we use the values given in Table \ref{tab:pmns} at the
  scale of right-handed neutrino masses, neglecting the
  renormalization group running of the neutrino mass and mixing
  parameters below the mass scale of the right-handed neutrinos.  At
  $M_t$, 2-loop and 3-loop SM thresholds are included.
\item In addition, we fix other input parameters:
  \begin{align}
    m_0, \ a_0,\ M_{1/2},\ \tan{\beta},\ {\rm sgn}(\mu),\
    \hat{\underline{M}}_{N},\ \hat{\Theta}_l,\
    \hat{\Theta}_M,\ O,\ H,
    \label{inputs}
  \end{align}
  where $M_{1/2}$ is SU(5) gaugino mass at the Planck scale,
  $\tan\beta$ is the ratio of the VEVs of up- and down-type Higgs
  bosons, and $\mu$ is supersymmetric Higgs mass parameter.  (Here, we
  assume that $\mu$ is real.)  For simplicity, we take $M_{1/2}=m_0$,
  $a_0=0$ and ${\rm sgn}(\mu)=+$. 
\item The SM parameters at the MSSM scale is obtained by using the SM
  renormalization group equations (RGEs).  In our analysis, the MSSM
  scale is define as the geometric mean of the stop mass eigenvalues,
  $M_S=\sqrt{m_{\tilde{t}_1}m_{\tilde{t}_2}}$; in our numerical
  calculation, $M_{\rm S}$ is determined iteratively (see the
  following arguments).  At $M_{\rm S}$, SUSY threshold corrections to
  the Higgs quartic coupling constant $\lambda$, gauge couplings and
  the top Yukawa coupling are included \cite{Bagnaschi:2014rsa}.
\item The MSSM parameters at the mass scale of the right-handed
  neutrino masses are obtained by using SOFTSUSY package
  \cite{Allanach:2001kg}.  The
  couplings in the MSSM and those of the MSSMNR are matched at the
  tree level.  Notice that each right-handed neutrino decouples from
  the effective theory at $\underline{M}_{N_i}$, and we use the
  effective theory without $N_i$ for the scale
  $Q<\underline{M}_{N_i}$; here, $N_i$ is defined in the basis in
  which it becomes the mass eigenstate of
  $M_N(Q=\underline{M}_{N_i})$.
\item In order to study the running between the mass scale of the
  right-handed neutrinos and the GUT scale, we modify SOFTSUSY package with
  including the coupling constants related to right-handed neutrinos.
  The RGEs of MSSMNR can be obtained in \cite{Masina:2003wt}.
\item The Yukawa matrices of the MSSMNR and those of the SU(5)NR are
  matched by using Eqs.\ \eqref{yuMgut} -- \eqref{MnMgut} at the GUT
  scale; in our numerical calculation, we take $M_{\rm{gut}}=2\times
  10^{16} {\rm GeV}$.  Other parameters are also matched accordingly.
  At the GUT scale, threshold corrections on down-type and
  charged-lepton type Yukawa matrices of MSSMNR are imposed following
  \cite{Evans:2018ewb} while tree level matching conditions are adopted for
  other dimensionless couplings and soft masses.
\item In order to take into account the running above the GUT scale,
  we also implement the RGEs in the SU(5)NR.  Here, we assume that
  the particle content of the SU(5)NR is minimal; $\Phi_i$, $\Psi_i$,
  $\Upsilon$, $H$, $\bar{H}$, and $\Sigma$, as well as the gauge
  multiplet.  The RGEs of the SU(5) gauge coupling constant and the
  gaugino mass are obtained based on this particle content.  In
  addition, for simplicity, we assume that the interactions of
  $\Sigma$ are so weak that their effects on the running are
  negligible.
\item The SUSY breaking parameters at the Planck scale are set by
  using the parameters $m_0$, $a_0$, and $M_{1/2}$.
\end{itemize}
For RGEs, we use 2-loop (SM and MSSM) and 1-loop (MSSMNR and SU(5)NR).
However, for the calculation of the Higgs quartic coupling constant,
we follow \cite{Bagnaschi:2014rsa} and solve 3-loop RGEs between $M_t$ and
$M_{\rm S}$.

With the modified SOFTSUSY package explained above, we calculate the MSSM
parameters at $Q=M_{\rm S}$ as follows.  For the consistency of
boundary conditions at $M_t$ and at $M_{\rm Pl}$, we iterate on $M_S$
and $M_H$.
\begin{enumerate}
\item We first fix the set of input parameters given in Eq.\ \eqref{inputs}.
  Then, we adopt the temporary values $M_S=m_0$ and
  $M_H=126\ \rm GeV$.
\item The boundary conditions for SM couplings at $Q=M_t$ are
  determined with following the procedure given in \cite{Buttazzo:2013uya},
  using $M_t$, Higgs mass $M_H$, $W$-boson mass $M_W$,
  $Z$-boson mass $M_Z$, the strong coupling constant
  $\alpha_s(M_Z)$ and fine structure constant $\alpha(M_Z)$
    as input parameters.
\item We solve the RGE runnings of the Yukawa coupling constants and
  the gauge coupling constants from $M_t$ to $M_{\rm Pl}$ and
  the Planck-scale values of those parameters are determined.
\item We set the boundary conditions for soft SUSY breaking parameters
  at $M_{\rm Pl}$ and run them down to $M_{\rm S}$. 
\item At $M_S$, the SM Higgs quartic coupling $\lambda$ is calculated
  from $\tan{\beta}$, with including the SUSY threshold
  corrections. The $\mu$-parameter and $B$-parameter are determined
  from the tree level electroweak symmetry breaking (EWSB) condition,
  which depends on $\tan{\beta}$ and Higgs soft masses.
  The Higgs quartic coupling at $Q=M_t$ is also determined.
\item Renew $M_S$ from stop masses and $M_H$ from SM couplings.
\item Iterate the steps from 2 to 6 until $M_S$ and $M_H$ converge.
\end{enumerate}

Once the MSSM parameters at $Q=M_{\rm S}$ are fixed, CP and LFV
observables are calculated by using those parameters.  After EWSB,
relevant operators of our interest are given by
\begin{align}
  {\cal L}^{\rm{eEDM}} &=
  -\frac{id_e}{2}\bar{\psi}_e\sigma^{\mu\nu}\gamma^5 \psi_e {F}_{\mu\nu},
  \\
  {\cal L}^{\mu\rightarrow e\gamma} &=
  -\frac{1}{2}\bar{\psi}_e\sigma^{\mu\nu}(a_LP_L+a_RP_R)\psi_{\mu} F_{\mu\nu},
  \label{muegamma}
\end{align}
where $F_{\mu\nu}$ is the field strength tensor of the photon,
$\psi_{\mu}$ and $\psi_e$ are field operators of muon and electron,
respectively, and $d_e$, $a_L$, and $a_R$ are coefficients. 
$d_e$ is the electron EDM while the decay rate of $\mu\rightarrow e\gamma$
process is give by
\begin{align}
\Gamma(\mu\rightarrow e\gamma)&=\frac{m_{\mu}^3}{16\pi}(|a_L|^2+|a_R|^2).
\end{align}
For the
LFV decay processes of $\tau$, like $\tau\rightarrow\mu\gamma$ and
$\tau\rightarrow e\gamma$, the operator is like that given in
Eq.\ \eqref{muegamma} with field operators being properly replaced.
For the detail about the calculation of the electron EDM and the LFV
decay rates, see, for example,
\cite{Hisano:1995cp,Hisano:2008hn}.

Before showing our numerical results, it is instructive to use the
leading-log approximation for the understandings of qualitative
behaviors.  Above $M_{\rm{GUT}}$, $f_u$ contributes to the
off-diagonal elements of the right-handed selectron mass matrix
$m_{\tilde{e}}$ and $f_{\nu}$ contribute to that of the left-handed
slepton mass matrix $m_{\tilde{l}}$. Below $M_{\rm{GUT}}$, there is no
extra violation production in $m_{\tilde{e}}$ from Yukawa interactions
but $m_{\tilde{l}}$ still acquires off-diagonal elements from
neutrino-type Yukawa interactions.  Assuming a universality of the
right-handed neutrino masses, the leading-log approximation gives
\begin{align}
  (m^2_{\tilde{l}})_{ij} & \simeq m_0^2 \delta_{ij}
  -\frac{1}{8\pi^2} (3m_0^2+a_0^2)
  (\hat{\Theta}_l^*\underline{y}_{\nu}^{\dagger}
  \underline{y}_{\nu}\hat{\Theta}_l)_{ij}
  \log\left(\frac{M_{\rm{Pl}}}{M_{N_R}}\right),
  \label{mtildel}\\
  (m^2_{\tilde{e}})_{ij} & \simeq m_0^2 \delta_{ij}
  -\frac{3}{8\pi^2}(3m_0^2+a_0^2)
  (V^T\hat{f}_{u}^2V^{*})_{ij}
  \log\left(\frac{M_{\rm{Pl}}}{M_{\rm{GUT}}}\right),
\end{align}
where $M_{N_R}$ is the universal right-handed neutrino mass. 
The off-diagonal elements
of $m^2_{\tilde{l}}$ are approximately proportional to
the corresponding elements of
$\hat{\Theta}_l^*{\underline{y}}_{\nu}^{\dagger}{\underline{y}}_{\nu}\hat{\Theta}_l$.
When the $r$ parameter is sizable we can find
\begin{align}
  \hat{\Theta}_l^*\underline{y}_{\nu}^{\dagger}{\underline{y}}_{\nu}\hat{\Theta}_l
  \simeq
  \frac{e^{2r}}{v_u^2}\mathrm{tr}\left(\tilde{P}(n)O^T \underline{\hat{M}}_N O\right)%\left(\tilde{n}^{\dagger}O^T \underline{\hat{M}}_N O \tilde{n}^{}\right)
  \tilde{U} \hat{m}_{\nu}^{\frac{1}{2}} \tilde{P}(n)% \tilde{n} \tilde{n}^{\dagger}
  \hat{m}_{\nu}^{\frac{1}{2}}\tilde{U}^{\dagger}
  + O(e^r),
  \label{yndgryn}
\end{align}
where $\tilde{U}\equiv\hat{\Theta}_l^*U_{\rm{PMNS}}\hat{\Theta}_M^*$.
The CI parameters may enhance the off diagonal elements of
$m^2_{\tilde{l}}$ because the magnitude of ${\underline{y}}_{\nu}^{\dagger}{\underline{y}}_{\nu}$
is proportional to $e^{2r}$.  From Eqs.\ \eqref{mtildel} and
\eqref{yndgryn}, one can see that the renormalization group effects on
$m^2_{\tilde{l}}$ is suppressed when the mass scale of the
right-handed neutrinos becomes smaller.  Thus, when the effects of the
CI parameters are neglected, the CP and flavor violations due to the
renormalization group effects are highly suppressed when the mass
scale of the right handed neutrinos is much smaller than $\sim
10^{14}\ {\rm GeV}$.  With the CI parameters, this may not be the
case.  One can see that, when the $r$ parameter is larger than $\sim
1$, the renormalization group effects can be sizable even when the
right-handed neutrinos are relatively light.  In the following, we
will see that the enhancement due to the CI parameters can indeed
enhance the electron EDM and LFV decay rates.

\begin{figure}[t]
\centering
\begin{minipage}{0.43\hsize}
\begin{tikzpicture}[baseline= {([yshift=+13.8ex]current bounding box.center)}]
\tikzset{gauge/.style={decorate,decoration={snake,amplitude=1,segment length=4}}}
\tikzset{forarrow/.style={postaction={decorate,decoration={markings,mark=at position 0.5 with {\arrow{>}}}},>=stealth}}
\tikzset{invarrow/.style={postaction={decorate,decoration={markings,mark=at position 0.5 with {\arrow{<}}}},>=stealth}}
\tikzset{intarrow/.style={postaction={decorate,decoration={markings,mark=at position 0.25 with {\arrow{>}},mark=at position 0.75 with {\arrow{<}}}},>=stealth}}
\tikzset{extarrow/.style={postaction={decorate,decoration={markings,mark=at position 0.75 with {\arrow{>}},mark=at position 0.25 with {\arrow{<}}}},>=stealth}}
%scalar
\coordinate (A) at (0,0)  node[below,scale=0.7]{$-\sqrt{2}g'$};
\draw[dashed,invarrow](A)--++(1.35,0)node[draw,fill=black,circle,inner sep=1.4]{} node[midway,above,scale=0.8]{$\tilde{e}_R$}node[below,scale=0.8]{$-(m^2_{\tilde{e}})_{1,3}$}coordinate(B);
\draw[invarrow,dashed](B)--++(1.35,0)node[draw,fill=black,circle,inner sep=1.4]{}  node[midway,above,scale=0.8]{$\tilde{\tau}_R$}node[below=1,scale=0.8]{$\mu\tan{\beta}$}node[below=12,scale=0.8]{$m_{\tau}$}coordinate(C);
\draw[dashed,invarrow](C)--++(1.35,0)node[draw,fill=black,circle,inner sep=1.4]{} node[midway,above,scale=0.8]{$\tilde{\tau}_L$}node[below,scale=0.8]{$-({m}^2_{\tilde{l}})_{3,1}$}coordinate(D) ;
\draw[dashed,invarrow](D)--++(1.35,0)node[midway,above,scale=0.8]{$\tilde{e}_L$}coordinate(E)node[below,scale=0.8]{$\frac{g'}{\sqrt{2}}$};
%fermion
\path (A)--++(2.7,2.7)coordinate(G);
\draw[invarrow](A) arc(180:90:2.7) (G)node[below,scale=0.8]{$\tilde{B}$};
\draw[forarrow](G) arc(90:0:2.7)(E);
%external
\draw[invarrow](A)--++(-0.8,0)node[midway,above,scale=0.8]{$\bar{e}$};
\draw[invarrow](E)--++(0.8,0)node[midway,above,scale=0.8]{$e$};
%gauge
\path (E)--++(0,1)coordinate(F);
\draw[gauge](F)--++(0.6,0.6)node[midway,right]{$$}; 
\end{tikzpicture}
\end{minipage}
\hspace{10mm}
\begin{minipage}{0.43\hsize}
\begin{tikzpicture}[baseline= {([yshift=0ex]current bounding box.center)}]
\tikzset{gauge/.style={decorate,decoration={snake,amplitude=1,segment length=4}}}
\tikzset{forarrow/.style={postaction={decorate,decoration={markings,mark=at position 0.5 with {\arrow{>}}}},>=stealth}}
\tikzset{invarrow/.style={postaction={decorate,decoration={markings,mark=at position 0.5 with {\arrow{<}}}},>=stealth}}
\tikzset{intarrow/.style={postaction={decorate,decoration={markings,mark=at position 0.25 with {\arrow{>}},mark=at position 0.75 with {\arrow{<}}}},>=stealth}}
\tikzset{extarrow/.style={postaction={decorate,decoration={markings,mark=at position 0.75 with {\arrow{>}},mark=at position 0.25 with {\arrow{<}}}},>=stealth}}
%scalar
\coordinate (A) at (0,0)  node[below,scale=0.7]{$-g$};
\draw[dashed,invarrow](A)--++(2.7,0)node[draw,fill=black,circle,inner sep=1.4]{} node[midway,above,scale=0.8]{$\tilde{\nu}_{eL}$}node[below,scale=0.8]{$-(m^2_{\tilde{l}})_{1,2}$}coordinate(B);
\draw[dashed,invarrow](B)--++(2.7,0)node[midway,above,scale=0.8]{$\tilde{\nu}_{\mu L}$}coordinate(E)node[below,scale=0.8]{$\frac{\sqrt{2}m_{\mu}}{v\cos{\beta}}$};
%fermion
\path (A)--++(2.7,2.7)coordinate(G);
\draw[intarrow](A) arc(180:90:2.7) node[pos=0.25,right,scale=0.7]{$\tilde{W}^{+}$}node[pos=0.75,below,scale=0.7]{$\tilde{W}^{-}$}node[draw,fill=black,circle,inner sep=1.4]{}node[above,scale=0.8]{$-gv\sin{\beta}/\sqrt{2}$};
\draw[intarrow](G) arc(90:0:2.7)(E)node[pos=0.25,below,scale=0.7]{$\tilde{h}^+_u$}node[pos=0.75,left,scale=0.7]{$\tilde{h}^-_d$};
%external
\draw[forarrow](A)--++(-0.8,0)node[midway,above,scale=0.8]{$e$};
\draw[forarrow](E)--++(0.8,0)node[midway,above,scale=0.8]{$\bar{\mu}$};
%gauge
\path (E)--++(0,1)coordinate(F);
\draw[gauge](F)--++(0.6,0.6)node[midway,right]{$$}; 
\end{tikzpicture}
\end{minipage}
\caption{Examples of the mass insertion contributions to $d_e$ (left) and to
  ${\rm{Br}}(\mu\rightarrow e\gamma))$ (right).}
\label{fig:diagrams}
\end{figure}
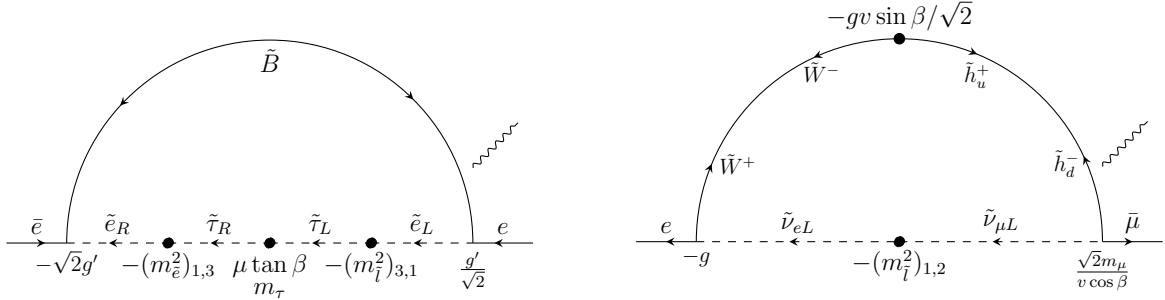

The off diagonal elements of the slepton mass matrix become the
sources of CP and flavor violations.  Although we numerically
calculate the electron EDM and LFV decay rates in the mass basis, with
which the effects of the off diagonal elements are taken into account
at all orders, it is suggestive to consider the mass insertion
method to understand the behaviors of the results.
Fig.\ \ref{fig:diagrams} shows the examples of the diagrams
contributing to the electron EDM and $\mu\rightarrow e\gamma$ process
in the mass insertion approximation.  In fact, when $\tan{\beta}\gg
1$, the dominant contributions to $\mu\rightarrow e\gamma$ originates
from a diagram with a mass insertion of $(m_{\tilde{l}})_{1,2}$.  For
the electron EDM, the dominant contribution is from a diagram with mass
insertions of $(m_{\tilde{e}})_{1,3}$ and $(m_{\tilde{l}})_{1,3}$, if
there's no CP phase in $\mu$ parameter (see
Fig.\ \ref{fig:diagrams}).  For the choice of parameters we adopt in
the following analysis, we found that the diagrams shown in
Fig.\ \ref{fig:diagrams} become dominant for the electron EDM and the
decay rate for the process $\mu\rightarrow e\gamma$.

Now, we show the results of our numerical calculations.  Unless
otherwise stated, we take $\tan{\beta}=8$ and $m_0=10\ \rm TeV$, which
give $M_H\simeq 126\ \rm GeV$.  We neglect the effects of Majorana
phases and simply set $\hat{\Theta}_M=\bm{1}$. The three
types of structures of $\underline{ \hat{M}}_N$ and $O$ are adopted:
\begin{itemize}
\item (U) Universal:
  \begin{align}
    O^{T}\underline{\hat{M}}_N O=M_{N_R}\bm{1}.
  \end{align}
\item (H) Hierarchical:
  \begin{align}
    O^{T}\underline{\hat{M}}_N O=M_{N_R}
    \left(
    \begin{array}{ccc}
      10^{-2} & 0 & 0 \\
      0 & 10^{-1} &0 \\
      0 & 0 & 1 
    \end{array}
    \right).
  \end{align}
\item (IH) Inverse hierarchical:
  \begin{align}
    O^{T}\underline{\hat{M}}_N O=M_{N_R}
    \left(
    \begin{array}{ccc}
      1 & 0 & 0 \\
      0 & 10^{-1} &0 \\
      0 & 0 & 10^{-2} 
    \end{array}
    \right).
  \end{align}
\end{itemize}

The Yukawa couplings may blow up if the CI parameter $r$ is too large.
In such a case, the perturbative calculation becomes unreliable.  In
order to avoid the blow up of the Yukawa couplings, we impose the
following constraints on the neutrino Yukawa couplings at any
renormalization scale:
\begin{align}
  \frac{\mathrm{tr}(f_{\nu}^{\dagger}f_{\nu})}{(4\pi)^2}<1,~~~
  \frac{\mathrm{tr}(y_{\nu}^{\dagger}y_{\nu})}{(4\pi)^2}<1.
\end{align}

\begin{figure}[t]
 \centering
    \begin{minipage}{0.44\hsize}
      \includegraphics[width=1\textwidth]{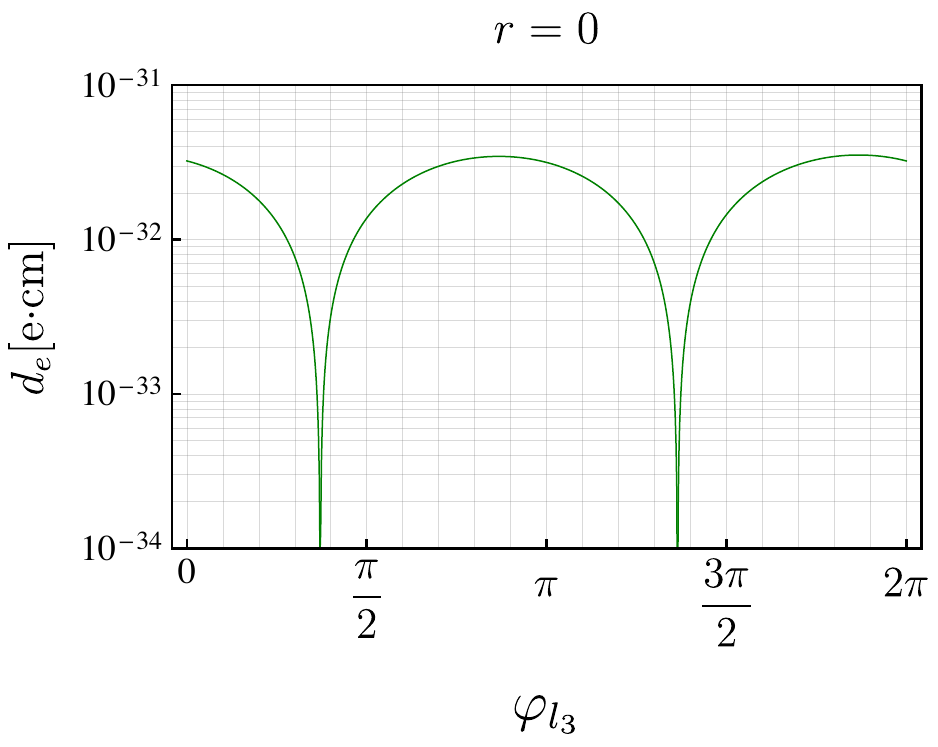}
    \end{minipage}
    \hspace{10mm}
    \begin{minipage}{0.44\hsize}
      \includegraphics[width=1\textwidth]{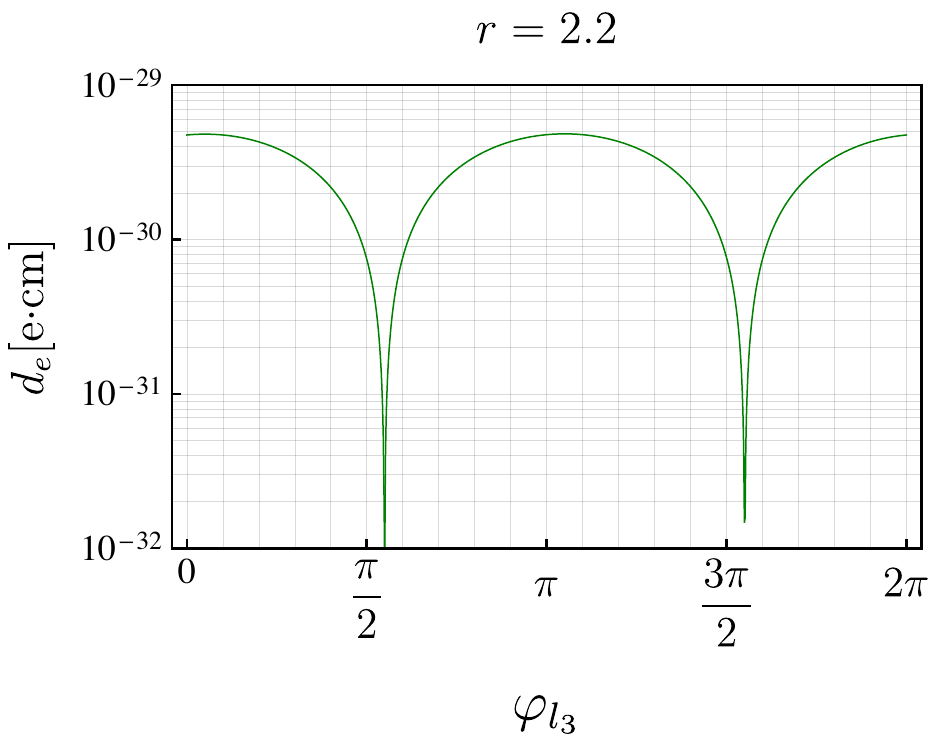}
    \end{minipage}
  \caption{The dependence of $d_e$ on GUT phase $\varphi_{l_{3}}$
    with $r=0$ (left) or $r=2.2$ (right). Here $\tan{\beta}=8$,
    $m_0=10\, \rm{TeV}$, $M_{N_R}=10^{13}\, \rm{GeV}$,  
    $(\theta,\phi)=(\frac{\pi}{2},0)$, $\hat{\Theta}_M=\bm{1}$
    and $\varphi_{l_{2}}=0$. }
  \label{fig:EdmPhase}
\end{figure}

In Fig.\ \ref{fig:EdmPhase}, we show how the electron EDM depends on
the GUT phases $\varphi_{l_{3}}$, adopting the structure (U) of
right-handed neutrino masses and $M_{N_R}=10^{13}\ \rm{GeV}$.  Here,
we take $R=\bm{1}$ (left) and $(r,\ \theta,\ \phi) = (2.2,\ \pi/2,\ 0)$
(right).  As shown in Fig.\ \ref{fig:EdmPhase}, the electron EDM is
sensitive to $\varphi_{l_{3}}$.  This is because, through the
renormalization group effects, $\varphi_{l_{3}}$ affects the complex
phase of $(m_{\tilde{l}})_{31}$, which the electron EDM is
(approximately) proportional to.  We can see that the position of the
peak is shifted with the introduction of the CI parameters because
they contain CP phases.  In the following analysis, in order to
(approximately) maximize the electron EDM, we tune the GUT phase
$\varphi_{l_{3}}$ so that the contribution of the mass insertion
diagram shown in Fig.\ \ref{fig:diagrams} (left), which gives the
dominant contribution to the electron EDM in most of the parameter
region in our study, is maximized.\footnote
{If $r$ is large enough, other mass-insertion diagrams with multiple
  insertions of $(m_{\tilde{l}})_{ij}$ become non-negligible. On the
  other hand, as $\underline{y}_{\nu}$ is small, the diagrams
  other than the left one of Fig.\ \ref{fig:diagrams} become
  sizable.}
We have also studied how the electron EDM depends on the phase
$\varphi_{l_{2}}$, and confirmed that such a dependence is weak.

\begin{figure}[t]
  \centering  
    \begin{minipage}{0.4\hsize}
      \includegraphics[width=1\textwidth]{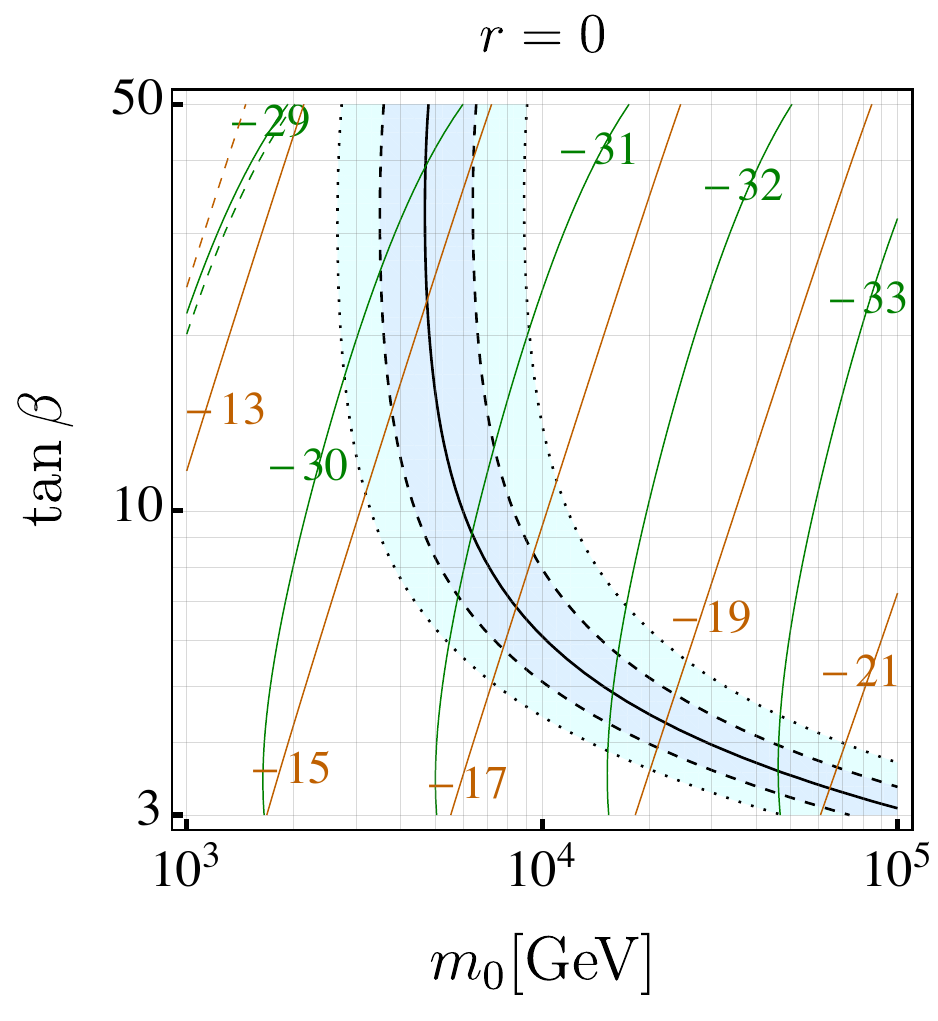}
    \end{minipage}
    \begin{minipage}{0.4\hsize}
      \includegraphics[width=1\textwidth]{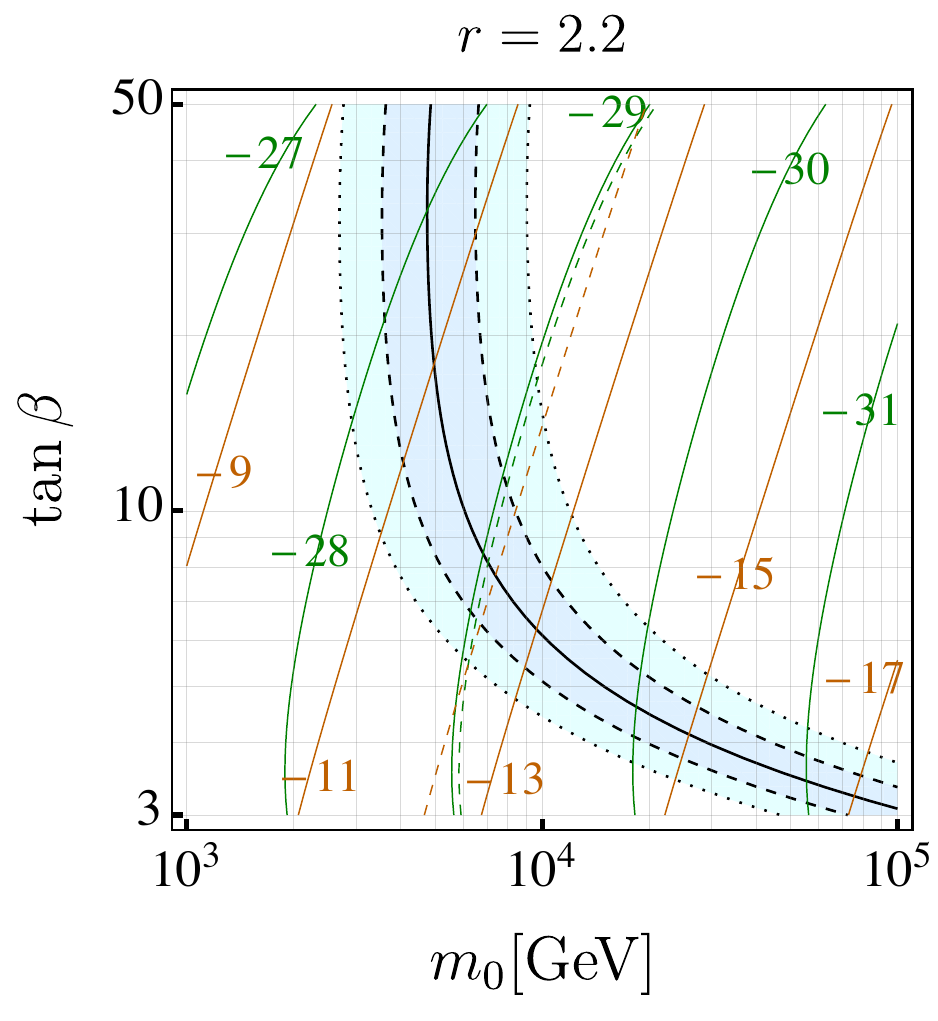}
    \end{minipage}
    \begin{minipage}{0.16\hsize}
      \includegraphics[width=1\textwidth]{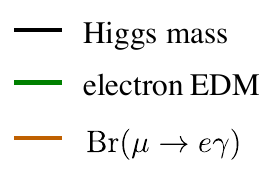}
    \end{minipage}
  \caption{Higgs mass (black), electron EDM (green) and
    Br($\mu\rightarrow e\gamma$) (orange) on $(m_0,\tan{\beta})$ plane
    with $r=0$ or $r=2.2$.  Here, we take $M_{N_R}=10^{13}\ \rm{GeV}$,
    $(\theta,\phi)=(\frac{\pi}{2},0)$, $\hat{\Theta}_M=\bm{1}$ and
    $\varphi_{l_{2}}=0$. $\varphi_{l_{3}}$ is chosen to maximize
    electron EDM.  The black lines are for
    $M_H=123,\ 124,\ 125,\ 126,\ 127\ \mathrm{GeV}$ from left to right.  The numbers
    in the figure are $\log_{10}d_e$ (green) and $\log_{10}{\rm
      Br}(\mu\rightarrow e\gamma)$ (orange).  The green and orange
    dashed lines are experimental bounds on the electron EDM and
    Br$(\mu\rightarrow e\gamma)$, respectively.}
  \label{fig:m0tanb}
\end{figure}

Fig.\ \ref{fig:m0tanb} shows the contours of constant Higgs
mass, maximized electron EDM and Br($\mu\rightarrow e\gamma$) on
$(m_0,\ \tan{\beta})$ plane, taking $M_{N_R}=10^{13}\, \rm{GeV}$.  In
the figure, we also show the contours on which $d_e$ and
Br($\mu\rightarrow e\gamma$) become equal to the current experimental
upper bounds; the upper bound on $d_e$ is given by ACME as
\cite{Andreev:2018ayy}
\begin{align}
  d_e<1.1 \times 10^{-29}\ e\, \mathrm{cm},
\end{align}
while the upper bound on the branching ratio for $\mu\rightarrow
e\gamma$ process is given by MEG experiment as \cite{TheMEG:2016wtm}
\begin{align}
  \mathrm{Br} (\mu\rightarrow e\gamma)<4.2\times 10^{-13}.
\end{align}
The left plot is for the case of $R={\bf 1}$, while the right one is
for the case of $(r,\ \theta,\ \phi) = (2.2,\ \pi/2,\ 0)$.  We see
that CI parameters have little influence on Higgs mass, but can
significantly enhance $d_e$ and Br($\mu\rightarrow e\gamma$).  This is
because, as we increase the $r$ parameter, the Yukawa couplings can
become larger (see Eq.\ \eqref{yndgryn}), which enhances the
renormalization group effects on $(m^2_{\tilde{l}})_{ij}$.  Because
the electron EDM and Br($\mu\rightarrow e\gamma$) are sensitive to the
off-diagonal elements of slepton mass squared matrices, the proper
introduction of the CI parameters has significant impact on the CP and
flavor violating observables.

\begin{figure}
  \centering
  \begin{minipage}{0.47\hsize}
    \includegraphics[width=1\textwidth]{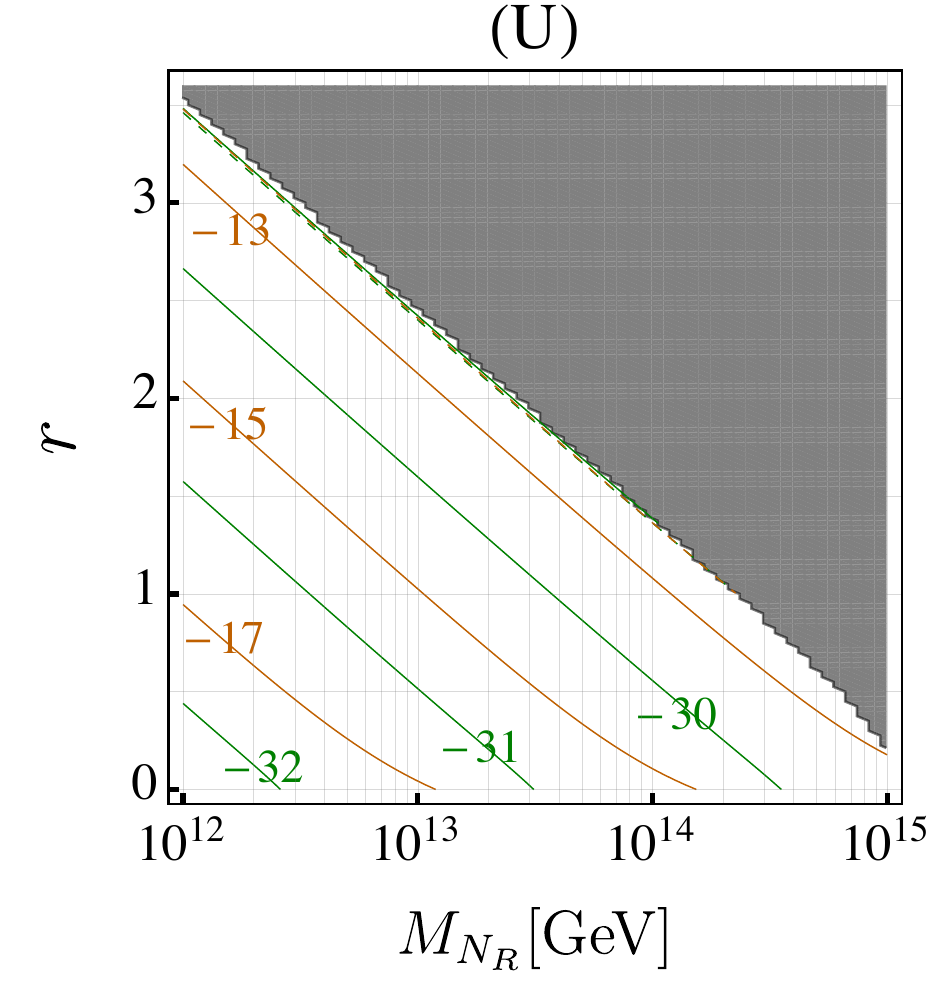}
  \end{minipage}
  \begin{minipage}{0.47\hsize}
    \includegraphics[width=1\textwidth]{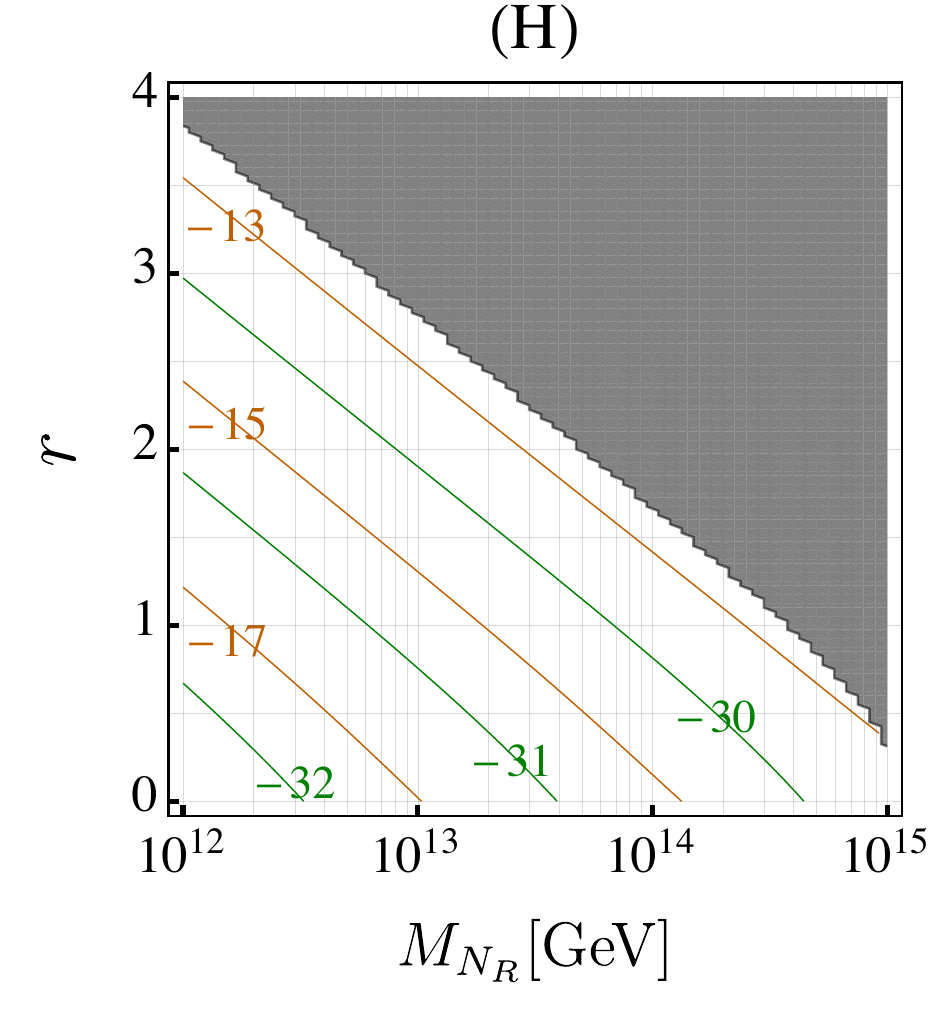}
  \end{minipage}\\
  \begin{minipage}{0.47\hsize}
    \includegraphics[width=1\textwidth]{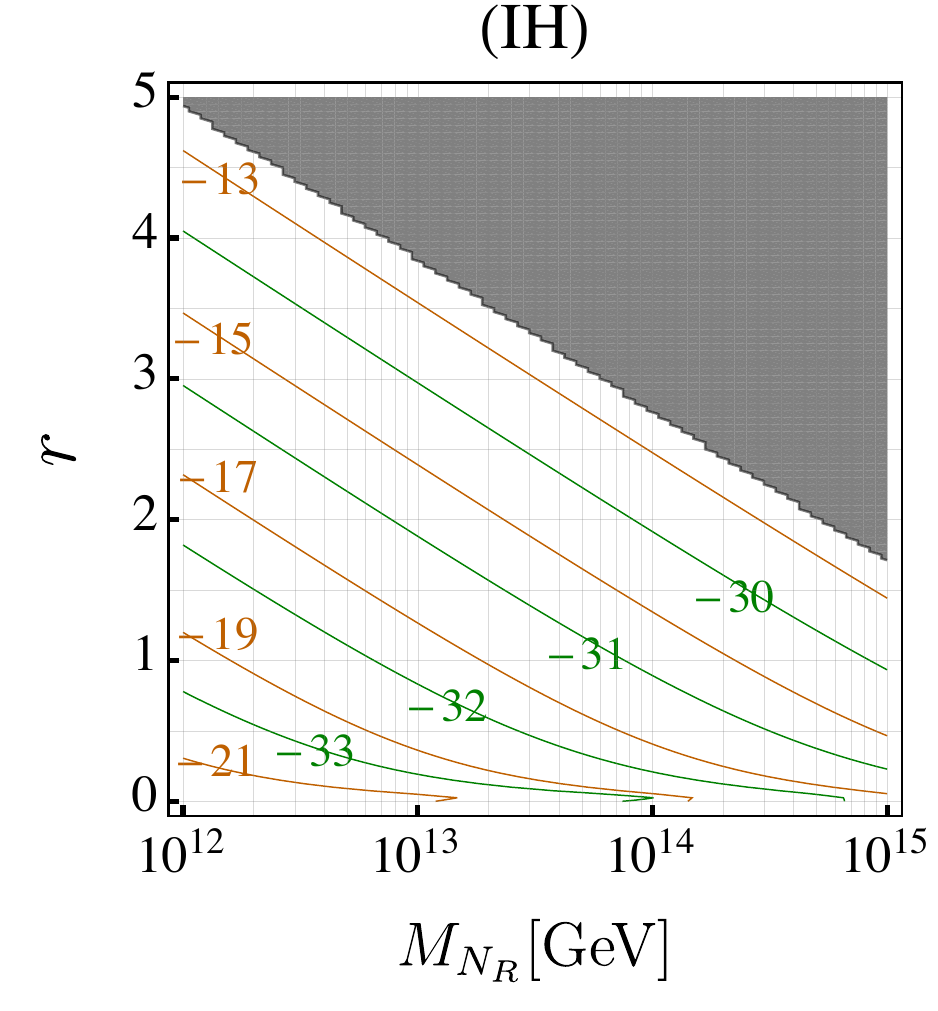}
   \end{minipage}
  \hspace{22mm}
  \begin{minipage}{0.20\hsize}
    \includegraphics[width=1\textwidth]{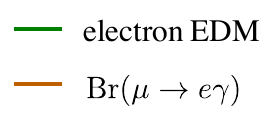}
  \end{minipage}
  \hspace{20mm}
  \caption{The electron EDM (green) and ${\rm{Br}} (\mu\rightarrow
    e\gamma)$ (orange) on $(M_{N_R},r)$ plane for the structures (U),
    (H) and (IH) of $O^T\underline{\hat{M}} _{N}O$.  We take
    $\tan{\beta}=8$, $m_0=10\, {\rm TeV}$, $(\theta,\ \phi)=(\pi/2,0)$
    , $\hat{\Theta}_M=\bm{1}$ and $\varphi_{l_{2}}=0$.  $\varphi_{l_{3}}$ is
    chosen to maximize $d_e$.  The numbers in the figure are
    $\log_{10}d_e$ (green) and $\log_{10}{\rm Br}(\mu\rightarrow
    e\gamma)$ (orange). }
  \label{fig:Mn-r}
\end{figure}

\begin{figure}
\centering
 \hspace{34mm}
  \begin{minipage}{0.47\hsize}
    \includegraphics[width=1\textwidth]{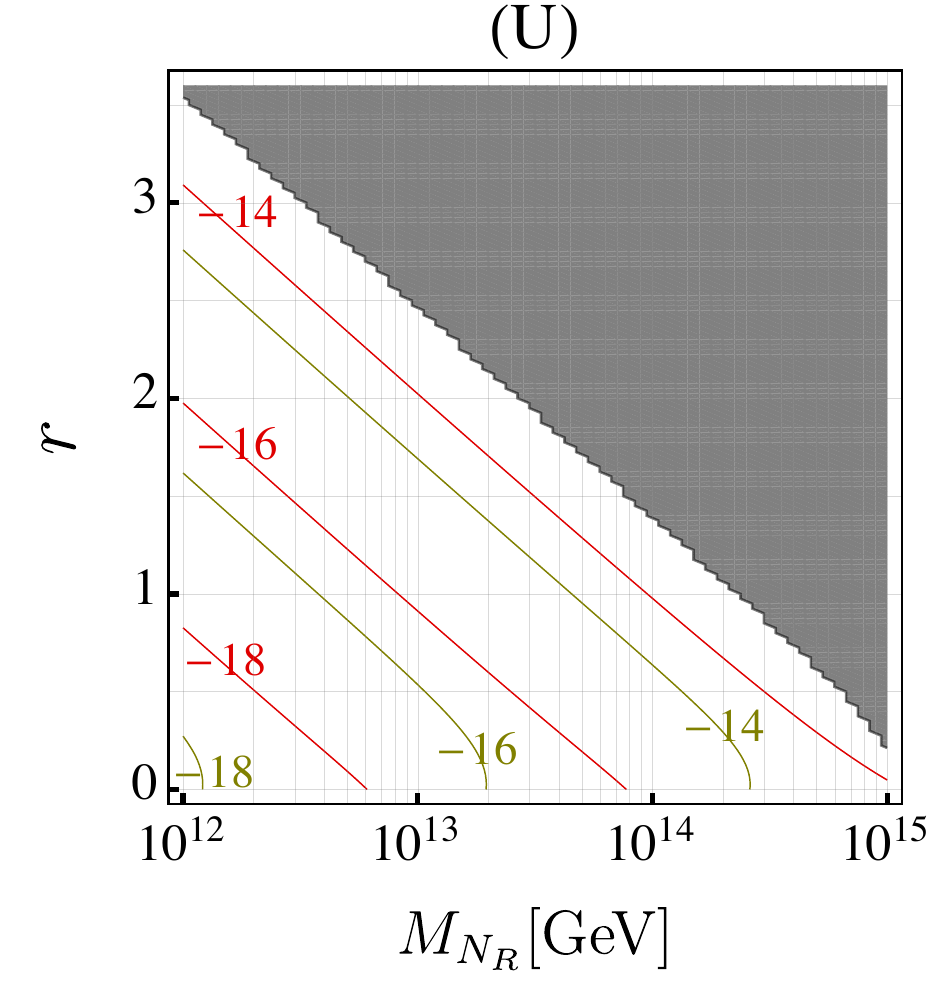}
  \end{minipage}
  \begin{minipage}{0.18\hsize}
    \includegraphics[width=1\textwidth]{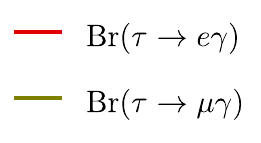}
  \end{minipage}
  \caption{ ${\rm{Br}}(\tau\rightarrow e\gamma)$ (red) and
    ${\rm{Br}}(\tau\rightarrow \mu\gamma)$ (yellow) on $(M_{N_R},r)$
    plane for the case (U) of universal right-handed neutrinos.  Other
    parameters are same as those in Fig.\ \ref{fig:Mn-r}.  The numbers
    are $\log_{10}{\rm{Br}}(\tau\rightarrow e\gamma)$ (red) and
    $\log_{10}{\rm{Br}}(\tau\rightarrow \mu\gamma)$ (yellow). }
  \label{fig:Mn-r2}
%\end{figure}
%%
%%
  %\begin{figure}[t]
  \centering  
    \begin{minipage}{0.47\hsize}
      \includegraphics[width=1\textwidth]{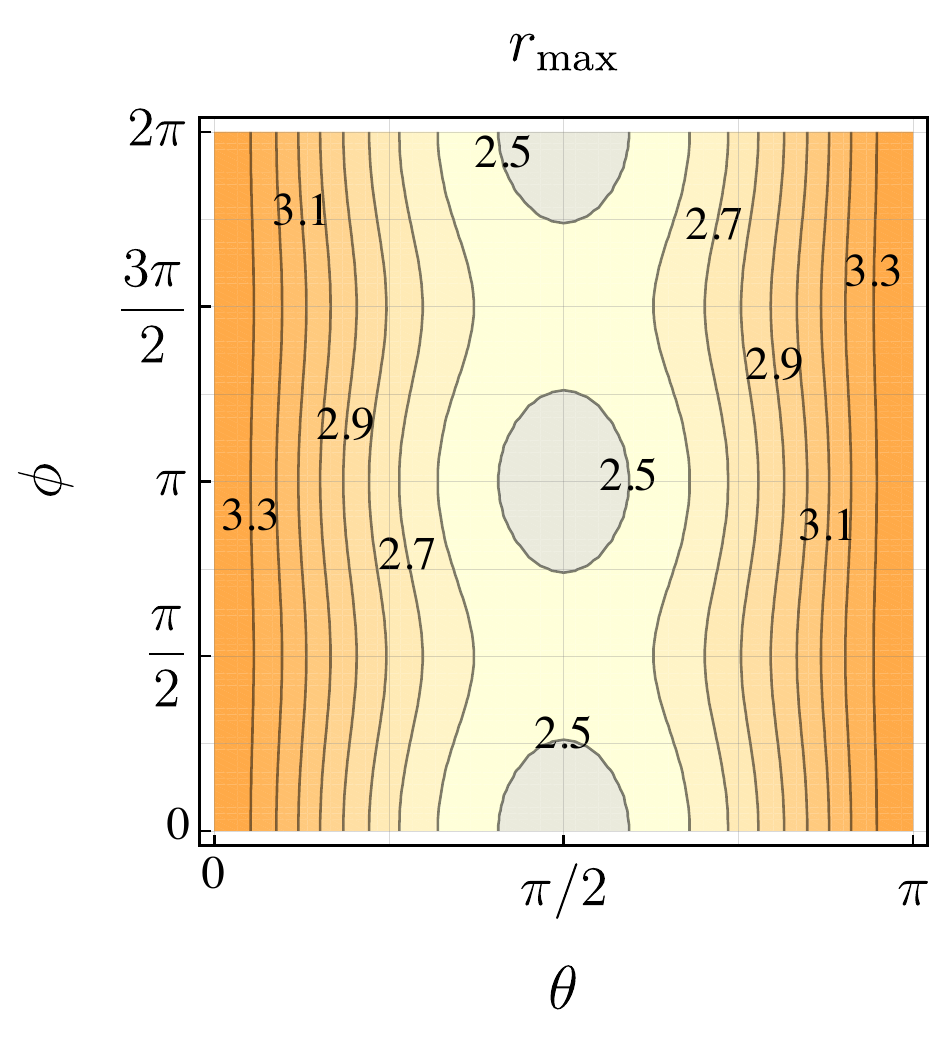}
    \end{minipage}
  \caption{The maximized $r$ parameter under the perturbativity
    constraints (upper-left). Here,
    right-handed neutrinos are (U) universal with $M_{N_R}=
    10^{13}\, \rm{GeV}$. $\tan{\beta}=8$, $m_0=10\, {\rm TeV}$,
    $\hat{\Theta}_M=\bm{1}$ and $\varphi_{l_{2}}=0$.  $\varphi_
    {l_{3}}$ is chosen to maximize electron EDM.
  }
  \label{fig:maxr}
\end{figure}

In Fig.\ \ref{fig:Mn-r}, we show contours of constant maximized
electron EDM and ${\rm Br}(\mu\rightarrow e\gamma)$ on $(M_{N_R},\ r)$
plane, taking $(\theta,\phi)=(\pi/2,0)$.  (In the figure, we shade
the regions in which the perturbativity of the Yukawa couplings breaks
down.)  Over the wide range of the parameter space, we have checked
that the mass-insertion diagrams shown in Fig.\ \ref{fig:diagrams} are
dominant.  We can see that the maximal possible values of the electron
EDM and ${\rm Br}(\mu\rightarrow e\gamma)$ are insensitive to the
scale and the structure of the right-handed neutrino mass matrix.
This is because the enhancement of the neutrino Yukawa couplings due
to the factor of $e^{2r}$ compensates the suppression due to the
smallness of the right-handed neutrino mass (see
Eq.\ \eqref{yndgryn}). Fig.\ \ref{fig:Mn-r2} shows the contours of
constant ${\rm Br}(\tau\rightarrow e\gamma)$ and ${\rm
  Br}(\tau\rightarrow\mu\gamma)$.  They are also enhanced by CI
parameters but are fairly below the current experimental bounds.

These figures show our main conclusion that the leptonic CP and
flavor violating signals through the renormalization group effects can
be sizable irrespective of the mass scale of right-handed neutrinos.
This is a contrast to the case without taking into account the effects
of the CI parameters; without the CI parameters, the neutrino Yukawa
coupling constants become tiny when the right-handed neutrinos are
much lighter than $\sim 10^{14}\, {\rm GeV}$.  In other words,
we have a chance to observe the leptonic CP and/or flavor violating
signals from the renormalization group effect even when the
right handed neutrino masses are relatively small.

\begin{figure}
  \centering
    \begin{minipage}{0.48\hsize}
      \includegraphics[width=1\textwidth]{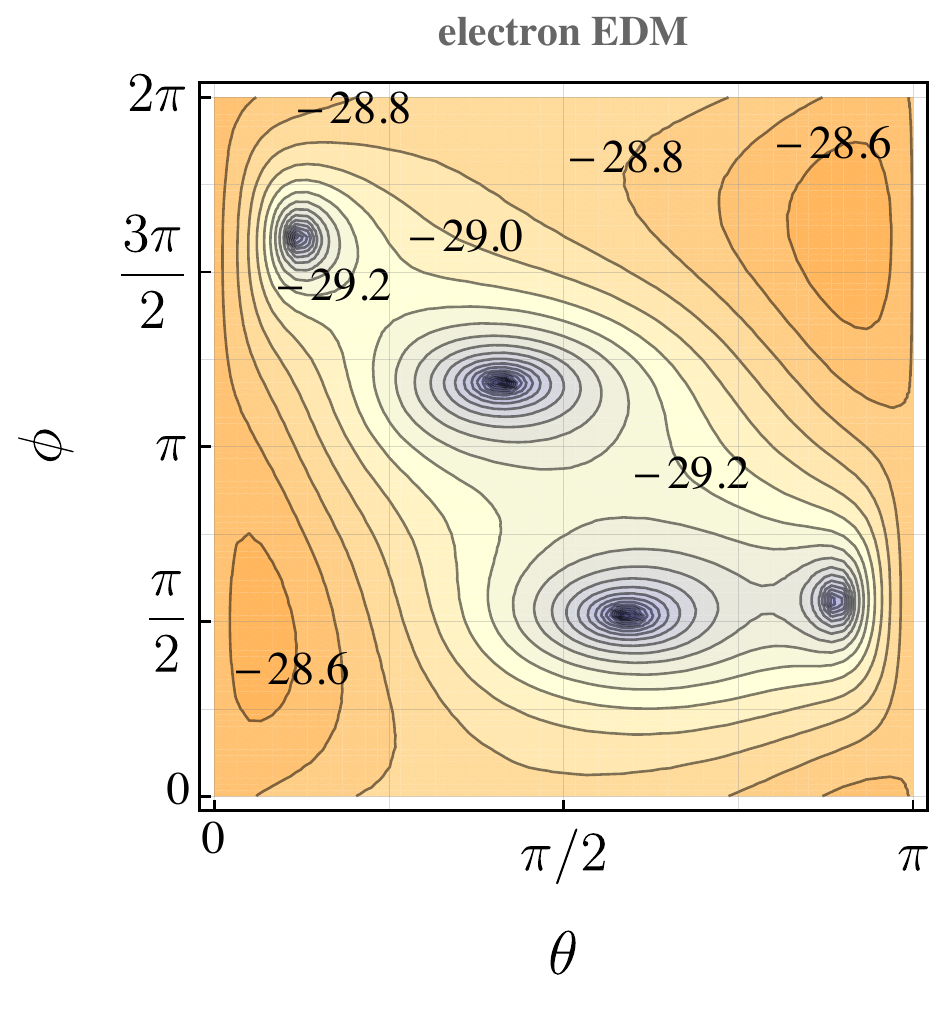}
    \end{minipage}
    \begin{minipage}{0.48\hsize}
      \includegraphics[width=1\textwidth]{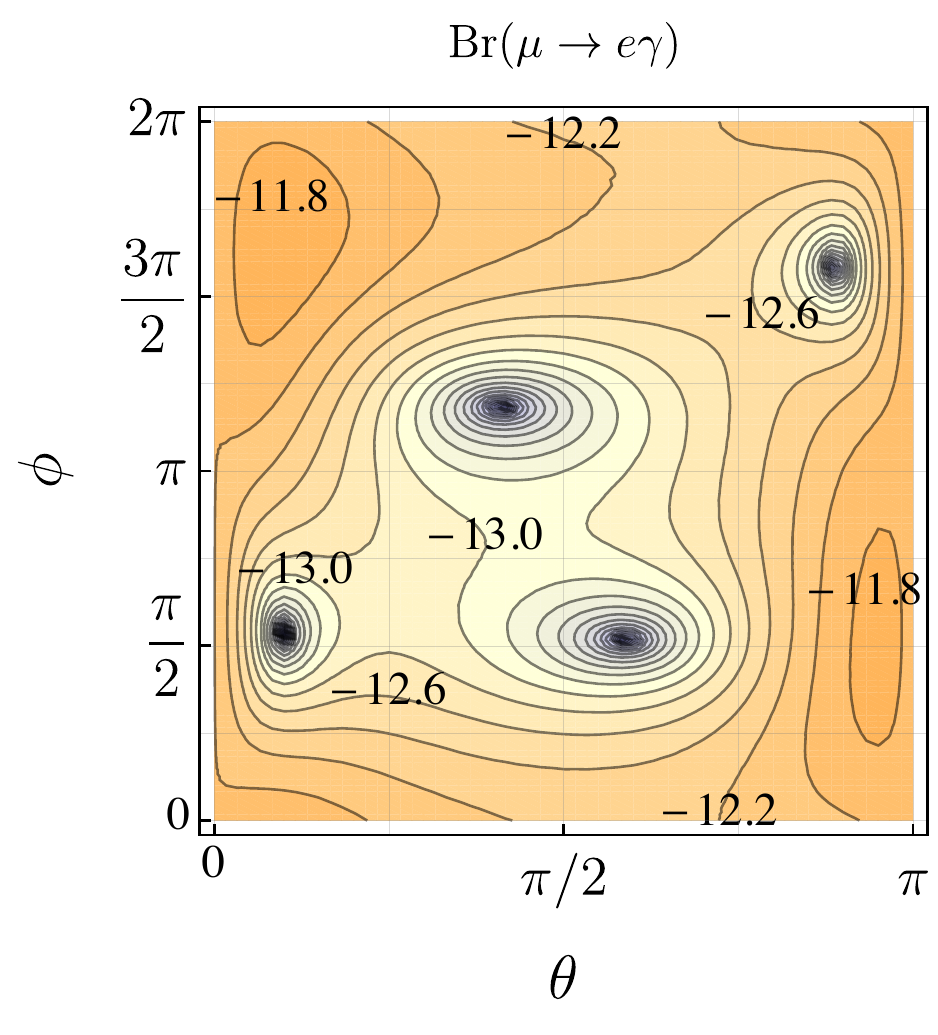}
    \end{minipage}\\
    \begin{minipage}{0.48\hsize}
      \includegraphics[width=1\textwidth]{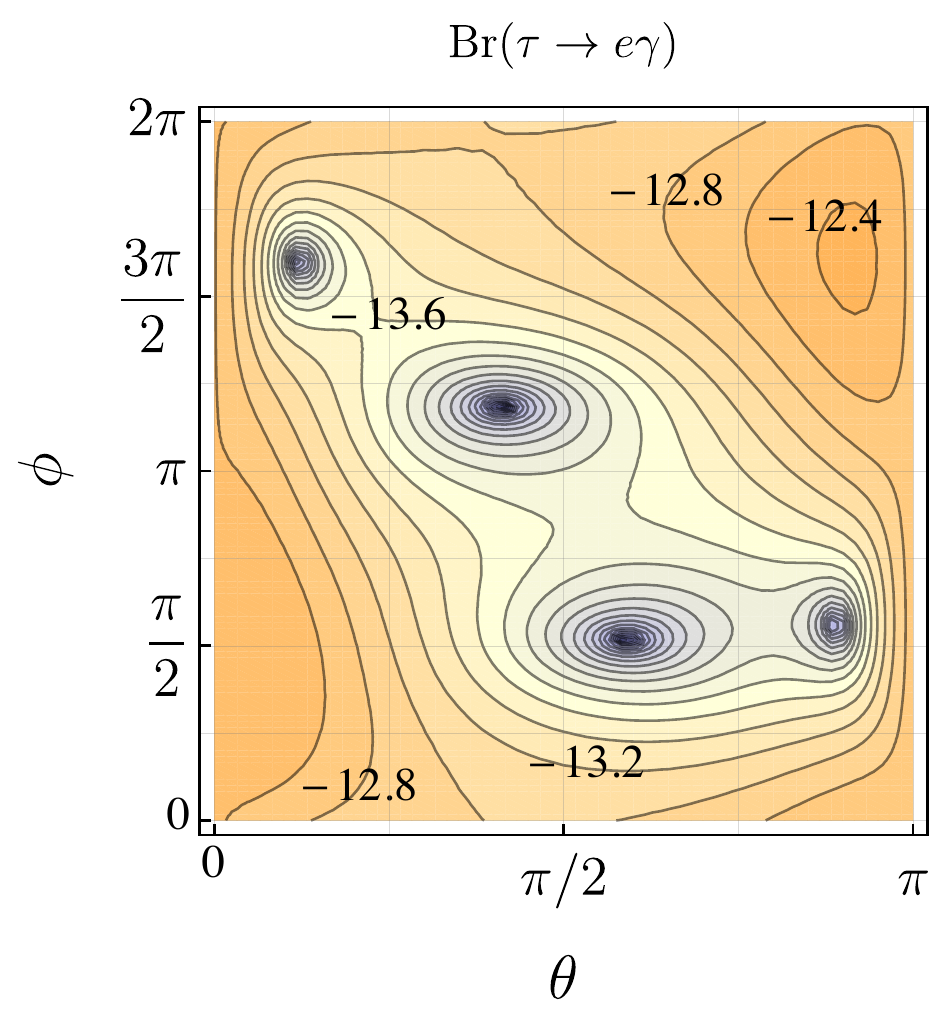}
    \end{minipage}
    \begin{minipage}{0.48\hsize}
      \includegraphics[width=1\textwidth]{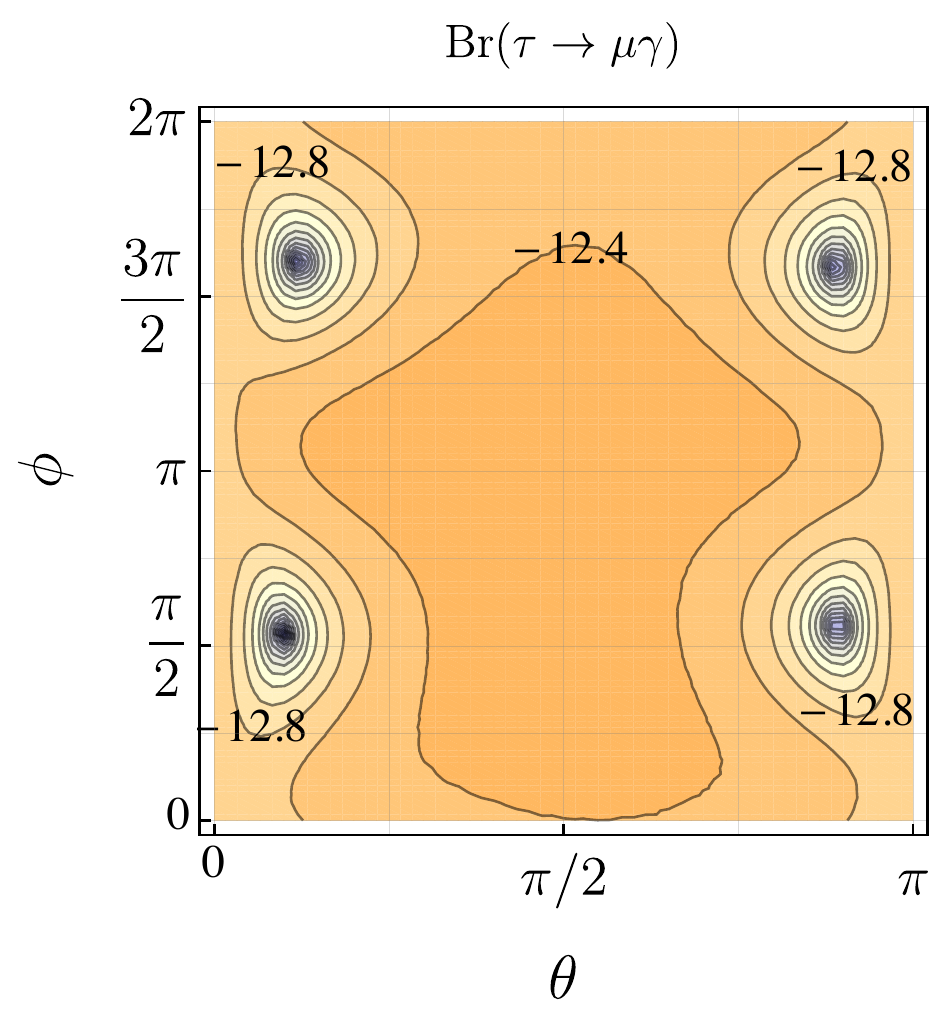}
    \end{minipage}
   \caption{Contours of constant electron EDM (upper-left),
    ${\rm{Br}}(\mu\rightarrow e\gamma)$ (upper-right),
    and
    ${\rm{Br}}(\tau\rightarrow l_j\gamma)$ (lower), calculated with
    the maximal possible value of $r$ shown in Fig.\ \ref{fig:maxr}.
    The model parameters are the same as those used in Fig.\ \ref{fig:maxr}.
  }
  \label{fig:th-ph}
\end{figure}

Fig.\ \ref{fig:maxr} shows how large the $r$ parameter can be on
$(\theta, \phi)$ plane, taking (U) universal right-handed neutrinos
with $M_{N_R}=10^{13}\, {\rm GeV}$.  With the choice of parameters
adopted in Fig.\ \ref{fig:maxr}, $r$ is required to be smaller than
about $2.5-3.3$.
Using the maximal possible value of $r$ given in 
Fig.\ \ref{fig:maxr}, we calculate the CP and flavor violating observables.
In Fig.\ \ref{fig:th-ph}, we show  maximized electron EDM,
${\rm{Br}}(\mu \rightarrow e\gamma)$, ${\rm{Br}}(\tau \rightarrow
e\gamma)$ and ${\rm{Br}}(\tau \rightarrow e\gamma)$.  We can see
that some of the observables are suppressed at particular points on
the $(\theta,\phi)$ plane and that the points of the suppressions are
correlated for different observables.  These are because, at the
points of the suppressions, two of $(m_{\tilde{l}})_{ij}$ (with $i\ne
j$) are simultaneously suppressed while the others are sizable.  This
can be understood as follows.  When $r$ is large, the off-diagonal
elements of $m_{\tilde{l}}$ can be approximated as
\begin{align}
  (m_{\tilde{l}})_{ij} \propto u_iu_j^*,
\end{align}
where (see Eq.\ \eqref{yndgryn})
\begin{align}
  u(\theta,\phi,\varphi_{M_k},\varphi_{l_k}) \equiv
  \hat{\Theta}_l^*U_{\rm{PMNS}}\hat{\Theta}_M^*
  \hat{m}^{\frac{1}{2}}_{\nu} \tilde{n}.
\end{align}
Thus, one of the elements $u_i$ becomes accidentally small,
$(m_{\tilde{l}})_{ij}$ ($j=1-3$) are all suppressed, resulting in the
correlation of the suppression points shown in Fig.\ \ref{fig:th-ph}.
For example, if $u_1$ is close to $0$, $(m_{\tilde{l}})_{1,2}$ and
$(m_{\tilde{l}})_{1,3}$, and hence $d_e$, ${\rm{Br}}(\mu
\rightarrow e\gamma)$ and ${\rm{Br}}(\tau \rightarrow e\gamma)$,
becomes simultaneously
suppressed; for the present choice of parameters, this happens
when 
$(\theta,\phi)\simeq (0.42\pi, 1.18\pi)$ and
$(0.58\pi, 0.52\pi)$.

\section{Conclusions and Discussion}
\label{sec:conclusion}
\setcounter{equation}{0}

We have studied the leptonic CP and flavor violating observables,
i.e., the electron EDM $d_e$ and the branching rations of lepton
flavor violating decays ${\rm{Br}}( l_i\rightarrow l_j\gamma)$, in the
minimal supersymmetric SU(5) GUT with three right-handed neutrinos. We
paid particular attention to the effects of the CI parameters
$R=OH(r,\theta,\phi)$ in the neutrino Yukawa matrix, which has not
been studied extensively before. With the assumption of the
universality boundary conditions for soft SUSY breaking masses, we
have calculated the electron EDM and ${\rm{Br}}( l_i\rightarrow
l_j\gamma)$ with varying CI parameters as well as the MSSM and GUT
parameters.  Imposing Higgs mass constraints as well as other
constraints from low-energy observations, we have studied how the CP
and flavor violating observables behaves.

In SUSY models, the off-diagonal elements of the slepton mass matrices
are induced by renormalization group effects in particular when there
exists right-handed neutrinos with sizable neutrino Yukawa couplings
or when quarks and leptons are unified into same multiplets of GUT.
The off-diagonal elements of the slepton mass matrices become sources
of the leptonic CP and flavor violating observables, i.e., the
electron EDM and ${\rm{Br}}( l_i\rightarrow l_j\gamma)$.  Without
taking into account the effects of the CI parameters, effects of the
right-handed neutrinos on the renormalization group runnings become
irrelevant if the mass scale of the right-handed neutrinos is small;
this is because, assuming the seesaw formula for the active neutrino
masses, the neutrino Yukawa coupling is suppressed as the right-handed
neutrino becomes lighter.  Effects of the CI parameters may compensate
such an effect, and we found that the maximal possible values of $d_e$
and ${\rm{Br}}( l_i\rightarrow l_j\gamma)$ are insensitive to the
structure of right-handed neutrino masses $\underline{\hat{M}}_N$ and
the orthogonal matrix $O$.  Especially, there are points where 2 of 3
independent off-diagonal elements are simultaneously suppressed.
Therefore, experimental studies of all the CP and flavor violating
observables are important to probe the model.

One interesting implication of our analysis should be on the
leptogenesis scenario \cite{Fukugita:1986hr}, in which the lepton
asymmetry generated by the decay of the right-handed neutrino is
converted to the baryon asymmetry of the universe.  In a simple
leptogenesis scenario, the mass scale of the lightest right-handed
neutrino is required to be larger than $\sim 10^{9-10}\, {\rm GeV}$
\cite{Giudice:2003jh, Buchmuller:2004nz}, while it should be smaller
than the reheating temperature after inflation in order not to dilute
the generated baryon asymmetry.  The total amount of the baryon
asymmetry generated by the leptogenesis scenario depends on the
detailed structure of the neutrino Yukawa couplings and neutrino mass
matrix.  The detailed analysis of the leptonic CP and flavor violating
observables in connection with the leptogenesis scenario is left for a
future work \cite{HiraoMoroiFuture}.

In this paper, we have concentrated on leptonic CP and flavor
violations.  In SUSY GUT with right-handed neutrinos, however, it is
also notable that sizable off-diagonal elements of squark mass
matrices may be also generated via the renormalization group effects.
In particular, above the GUT scale, the neutrino Yukawa interactions
affect the renormalization group runnings of the left-handed sdown
mass matrix; such an effect should also be sensitive to the CI
parameters.  The renormalization group effects on the squark mass
matrices, as well as hadronic CP and flavor violations in connection
with such effects, will be studied elsewhere \cite{HiraoMoroiFuture}.

\section*{Acknowledgment}

The work of TM is supported by JSPS KAKENHI Grant Nos.\
16H06490 and 18K03608.


\begin{thebibliography}{99}


%\cite{Zyla:2020zbs}
\bibitem{Zyla:2020zbs}
P.~A.~Zyla \textit{et al.} [Particle Data Group],
``Review of Particle Physics,''
PTEP \textbf{2020}, no.8, 083C01 (2020)
%doi:10.1093/ptep/ptaa104
%740 citations counted in INSPIRE as of 06 Feb 2021

%\cite{Okada:1990vk}
\bibitem{Okada:1990vk}
Y.~Okada, M.~Yamaguchi and T.~Yanagida,
``Upper bound of the lightest Higgs boson mass in the minimal supersymmetric standard model,''
Prog. Theor. Phys. \textbf{85}, 1-6 (1991)
%doi:10.1143/ptp/85.1.1
%1396 citations counted in INSPIRE as of 06 Feb 2021

%\cite{Okada:1990gg}
\bibitem{Okada:1990gg}
Y.~Okada, M.~Yamaguchi and T.~Yanagida,
``Renormalization group analysis on the Higgs mass in the softly broken supersymmetric standard model,''
Phys. Lett. B \textbf{262}, 54-58 (1991)
%doi:10.1016/0370-2693(91)90642-4
%628 citations counted in INSPIRE as of 06 Feb 2021

%\cite{Ellis:1990nz}
\bibitem{Ellis:1990nz}
J.~R.~Ellis, G.~Ridolfi and F.~Zwirner,
``Radiative corrections to the masses of supersymmetric Higgs bosons,''
Phys. Lett. B \textbf{257}, 83-91 (1991)
%doi:10.1016/0370-2693(91)90863-L
%1554 citations counted in INSPIRE as of 06 Feb 2021

%\cite{Haber:1990aw}
\bibitem{Haber:1990aw}
H.~E.~Haber and R.~Hempfling,
``Can the mass of the lightest Higgs boson of the minimal supersymmetric model be larger than m(Z)?,''
Phys. Rev. Lett. \textbf{66}, 1815-1818 (1991)
%doi:10.1103/PhysRevLett.66.1815
%1550 citations counted in INSPIRE as of 06 Feb 2021

\bibitem{Wells:2004di}
J.~D.~Wells,
%``PeV-scale supersymmetry,''
Phys. Rev. D \textbf{71}, 015013 (2005)
[arXiv:hep-ph/0411041 [hep-ph]].
%305 citations counted in INSPIRE as of 07 Feb 2021

\bibitem{Giudice:1998xp}
G.~F.~Giudice, M.~A.~Luty, H.~Murayama and R.~Rattazzi,
%``Gaugino mass without singlets,''
JHEP \textbf{12}, 027 (1998)
%doi:10.1088/1126-6708/1998/12/027
[arXiv:hep-ph/9810442 [hep-ph]].

\bibitem{Ibe:2006de}
M.~Ibe, T.~Moroi and T.~T.~Yanagida,
%``Possible Signals of Wino LSP at the Large Hadron Collider,''
Phys. Lett. B \textbf{644}, 355-360 (2007)
%doi:10.1016/j.physletb.2006.11.061
[arXiv:hep-ph/0610277 [hep-ph]].

\bibitem{Ibe:2011aa}
M.~Ibe and T.~T.~Yanagida,
%``The Lightest Higgs Boson Mass in Pure Gravity Mediation Model,''
Phys. Lett. B \textbf{709}, 374-380 (2012)
%doi:10.1016/j.physletb.2012.02.034
[arXiv:1112.2462 [hep-ph]].

\bibitem{ArkaniHamed:2012gw}
N.~Arkani-Hamed, A.~Gupta, D.~E.~Kaplan, N.~Weiner and T.~Zorawski,
%``Simply Unnatural Supersymmetry,''
[arXiv:1212.6971 [hep-ph]].

%\cite{Minkowski:1977sc}
\bibitem{Minkowski:1977sc}
P.~Minkowski,
``$\mu \to e\gamma$ at a Rate of One Out of $10^{9}$ Muon Decays?,''
Phys. Lett. B \textbf{67}, 421-428 (1977)
%doi:10.1016/0370-2693(77)90435-X
%4010 citations counted in INSPIRE as of 06 Feb 2021

%\cite{Yanagida:1979as}
\bibitem{Yanagida:1979as}
T.~Yanagida,
``Horizontal gauge symmetry and masses of neutrinos,''
Conf. Proc. C \textbf{7902131}, 95-99 (1979)
KEK-79-18-95.
%1890 citations counted in INSPIRE as of 06 Feb 2021

%\cite{GellMann:1980vs}
\bibitem{GellMann:1980vs}
M.~Gell-Mann, P.~Ramond and R.~Slansky,
``Complex Spinors and Unified Theories,''
Conf. Proc. C \textbf{790927}, 315-321 (1979)
[arXiv:1306.4669 [hep-th]].
%3288 citations counted in INSPIRE as of 06 Feb 2021

%\cite{Fukugita:1986hr}
\bibitem{Fukugita:1986hr}
M.~Fukugita and T.~Yanagida,
``Baryogenesis Without Grand Unification,''
Phys. Lett. B \textbf{174}, 45-47 (1986)
%doi:10.1016/0370-2693(86)91126-3
%3609 citations counted in INSPIRE as of 06 Feb 2021

%\cite{Borzumati:1986qx}
\bibitem{Borzumati:1986qx}
F.~Borzumati and A.~Masiero,
``Large Muon and electron Number Violations in Supergravity Theories,''
Phys. Rev. Lett. \textbf{57}, 961 (1986)
%doi:10.1103/PhysRevLett.57.961
%698 citations counted in INSPIRE as of 06 Feb 2021

%\cite{Barbieri:1994pv}
\bibitem{Barbieri:1994pv}
R.~Barbieri and L.~J.~Hall,
``Signals for supersymmetric unification,''
Phys. Lett. B \textbf{338}, 212-218 (1994)
%doi:10.1016/0370-2693(94)91368-4
[arXiv:hep-ph/9408406 [hep-ph]].
%474 citations counted in INSPIRE as of 06 Feb 2021

%\cite{Barbieri:1995tw}
\bibitem{Barbieri:1995tw}
R.~Barbieri, L.~J.~Hall and A.~Strumia,
``Violations of lepton flavor and CP in supersymmetric unified theories,''
Nucl. Phys. B \textbf{445}, 219-251 (1995)
%doi:10.1016/0550-3213(95)00208-A
[arXiv:hep-ph/9501334 [hep-ph]].
%559 citations counted in INSPIRE as of 06 Feb 2021

%\cite{Romanino:1996cn}
\bibitem{Romanino:1996cn}
A.~Romanino and A.~Strumia,
``Electric dipole moments from Yukawa phases in supersymmetric theories,''
Nucl. Phys. B \textbf{490}, 3-18 (1997)
%doi:10.1016/S0550-3213(97)00060-6
[arXiv:hep-ph/9610485 [hep-ph]].
%42 citations counted in INSPIRE as of 06 Feb 2021

%\cite{Hisano:1995nq}
\bibitem{Hisano:1995nq}
J.~Hisano, T.~Moroi, K.~Tobe, M.~Yamaguchi and T.~Yanagida,
``Lepton flavor violation in the supersymmetric standard model with seesaw induced neutrino masses,''
Phys. Lett. B \textbf{357}, 579-587 (1995)
%doi:10.1016/0370-2693(95)00954-J
[arXiv:hep-ph/9501407 [hep-ph]].
%408 citations counted in INSPIRE as of 06 Feb 2021

%\cite{Hisano:1995cp}
\bibitem{Hisano:1995cp}
J.~Hisano, T.~Moroi, K.~Tobe and M.~Yamaguchi,
``Lepton flavor violation via right-handed neutrino Yukawa couplings in supersymmetric standard model,''
Phys. Rev. D \textbf{53}, 2442-2459 (1996)
%doi:10.1103/PhysRevD.53.2442
[arXiv:hep-ph/9510309 [hep-ph]].
%725 citations counted in INSPIRE as of 06 Feb 2021

%\cite{Hisano:1998fj}
\bibitem{Hisano:1998fj}
J.~Hisano and D.~Nomura,
``Solar and atmospheric neutrino oscillations and lepton flavor violation in supersymmetric models with the right-handed neutrinos,''
Phys. Rev. D \textbf{59}, 116005 (1999)
%doi:10.1103/PhysRevD.59.116005
[arXiv:hep-ph/9810479 [hep-ph]].
%415 citations counted in INSPIRE as of 06 Feb 2021

%\cite{Baek:2000sj}
\bibitem{Baek:2000sj}
S.~Baek, T.~Goto, Y.~Okada and K.~i.~Okumura,
``Neutrino oscillation, SUSY GUT and B decay,''
Phys. Rev. D \textbf{63}, 051701 (2001)
%doi:10.1103/PhysRevD.63.051701
[arXiv:hep-ph/0002141 [hep-ph]].
%68 citations counted in INSPIRE as of 06 Feb 2021

%\cite{Moroi:2000mr}
\bibitem{Moroi:2000mr}
T.~Moroi,
``Effects of the right-handed neutrinos on Delta S = 2 and Delta B = 2 processes in supersymmetric SU(5) model,''
JHEP \textbf{03}, 019 (2000)
%doi:10.1088/1126-6708/2000/03/019
[arXiv:hep-ph/0002208 [hep-ph]].
%64 citations counted in INSPIRE as of 06 Feb 2021
 
%\cite{Moroi:2000tk}
\bibitem{Moroi:2000tk}
T.~Moroi,
``CP violation in B(d) ---\ensuremath{>} phi K(S)in SUSY GUT with right-handed neutrinos,''
Phys. Lett. B \textbf{493}, 366-374 (2000)
%doi:10.1016/S0370-2693(00)01160-6
[arXiv:hep-ph/0007328 [hep-ph]].
%155 citations counted in INSPIRE as of 06 Feb 2021

%\cite{Casas:2001sr}
\bibitem{Casas:2001sr}
J.~A.~Casas and A.~Ibarra,
``Oscillating neutrinos and $\mu \to e, \gamma$,''
Nucl. Phys. B \textbf{618}, 171-204 (2001)
%doi:10.1016/S0550-3213(01)00475-8
[arXiv:hep-ph/0103065 [hep-ph]].
%1100 citations counted in INSPIRE as of 06 Feb 2021

%\cite{Akama:2001em}
\bibitem{Akama:2001em}
N.~Akama, Y.~Kiyo, S.~Komine and T.~Moroi,
``CP violation in kaon system in supersymmetric SU(5) model with seesaw induced neutrino masses,''
Phys. Rev. D \textbf{64}, 095012 (2001)
%doi:10.1103/PhysRevD.64.095012
[arXiv:hep-ph/0104263 [hep-ph]].
%37 citations counted in INSPIRE as of 06 Feb 2021

%\cite{Ellis:2001xt}
\bibitem{Ellis:2001xt}
J.~R.~Ellis, J.~Hisano, S.~Lola and M.~Raidal,
``CP violation in the minimal supersymmetric seesaw model,''
Nucl. Phys. B \textbf{621}, 208-234 (2002)
%doi:10.1016/S0550-3213(01)00583-1
[arXiv:hep-ph/0109125 [hep-ph]].
%134 citations counted in INSPIRE as of 06 Feb 2021

%\cite{Ellis:2001yza}
\bibitem{Ellis:2001yza}
J.~R.~Ellis, J.~Hisano, M.~Raidal and Y.~Shimizu,
``Lepton electric dipole moments in nondegenerate supersymmetric seesaw models,''
Phys. Lett. B \textbf{528}, 86-96 (2002)
%doi:10.1016/S0370-2693(02)01197-8
[arXiv:hep-ph/0111324 [hep-ph]].
%96 citations counted in INSPIRE as of 06 Feb 2021

%\cite{Chang:2002mq}
\bibitem{Chang:2002mq}
D.~Chang, A.~Masiero and H.~Murayama,
``Neutrino mixing and large CP violation in B physics,''
Phys. Rev. D \textbf{67}, 075013 (2003)
%doi:10.1103/PhysRevD.67.075013
[arXiv:hep-ph/0205111 [hep-ph]].
%193 citations counted in INSPIRE as of 06 Feb 2021

%\cite{Ellis:2002fe}
\bibitem{Ellis:2002fe}
J.~R.~Ellis, J.~Hisano, M.~Raidal and Y.~Shimizu,
``A New parametrization of the seesaw mechanism and applications in supersymmetric models,''
Phys. Rev. D \textbf{66}, 115013 (2002)
%doi:10.1103/PhysRevD.66.115013
[arXiv:hep-ph/0206110 [hep-ph]].
%243 citations counted in INSPIRE as of 06 Feb 2021

%\cite{Hisano:2003bd}
\bibitem{Hisano:2003bd}
J.~Hisano and Y.~Shimizu,
``GUT relation in neutrino induced flavor physics in SUSY SU(5) GUT,''
Phys. Lett. B \textbf{565}, 183-192 (2003)
%doi:10.1016/S0370-2693(03)00638-5
[arXiv:hep-ph/0303071 [hep-ph]].
%64 citations counted in INSPIRE as of 06 Feb 2021

%\cite{Masina:2003wt}
\bibitem{Masina:2003wt}
I.~Masina,
``Lepton electric dipole moments from heavy states Yukawa couplings,''
Nucl. Phys. B \textbf{671}, 432-458 (2003)
%doi:10.1016/j.nuclphysb.2003.08.018
[arXiv:hep-ph/0304299 [hep-ph]].
%46 citations counted in INSPIRE as of 06 Feb 2021

%\cite{Ciuchini:2003rg}
\bibitem{Ciuchini:2003rg}
M.~Ciuchini, A.~Masiero, L.~Silvestrini, S.~K.~Vempati and O.~Vives,
``Grand unification of quark and lepton FCNCs,''
Phys. Rev. Lett. \textbf{92}, 071801 (2004)
%doi:10.1103/PhysRevLett.92.071801
[arXiv:hep-ph/0307191 [hep-ph]].
%72 citations counted in INSPIRE as of 06 Feb 2021

%\cite{Hisano:2004pw}
\bibitem{Hisano:2004pw}
J.~Hisano, M.~Kakizaki, M.~Nagai and Y.~Shimizu,
``Hadronic EDMs in SUSY SU(5) GUTs with right-handed neutrinos,''
Phys. Lett. B \textbf{604}, 216-224 (2004)
%doi:10.1016/j.physletb.2004.10.058
[arXiv:hep-ph/0407169 [hep-ph]].
%34 citations counted in INSPIRE as of 06 Feb 2021

%\cite{Calibbi:2006nq}
\bibitem{Calibbi:2006nq}
L.~Calibbi, A.~Faccia, A.~Masiero and S.~K.~Vempati,
``Lepton flavour violation from SUSY-GUTs: Where do we stand for MEG, PRISM/PRIME and a super flavour factory,''
Phys. Rev. D \textbf{74}, 116002 (2006)
%doi:10.1103/PhysRevD.74.116002
[arXiv:hep-ph/0605139 [hep-ph]].
%155 citations counted in INSPIRE as of 06 Feb 2021

%\cite{Hisano:2008df}
\bibitem{Hisano:2008df}
J.~Hisano and Y.~Shimizu,
``CP Violation in $B_s$ Mixing in the SUSY SU(5) GUT with Right-handed Neutrinos,''
Phys. Lett. B \textbf{669}, 301-305 (2008)
%doi:10.1016/j.physletb.2008.10.011
[arXiv:0805.3327 [hep-ph]].
%20 citations counted in INSPIRE as of 06 Feb 2021

%\cite{Hisano:2008hn}
\bibitem{Hisano:2008hn}
J.~Hisano, M.~Nagai and P.~Paradisi,
``Flavor effects on the electric dipole moments in supersymmetric theories: A beyond leading order analysis,''
Phys. Rev. D \textbf{80}, 095014 (2009)
%doi:10.1103/PhysRevD.80.095014
[arXiv:0812.4283 [hep-ph]].
%71 citations counted in INSPIRE as of 06 Feb 2021

%\cite{Borzumati:2009hu}
\bibitem{Borzumati:2009hu}
F.~Borzumati and T.~Yamashita,
``Minimal supersymmetric SU(5) model with nonrenormalizable operators: Seesaw mechanism and violation of flavour and CP,''
Prog. Theor. Phys. \textbf{124}, 761-868 (2010)
%doi:10.1143/PTP.124.761
[arXiv:0903.2793 [hep-ph]].
%51 citations counted in INSPIRE as of 06 Feb 2021

%\cite{Moroi:2013sfa}
\bibitem{Moroi:2013sfa}
T.~Moroi and M.~Nagai,
``Probing Supersymmetric Model with Heavy Sfermions Using Leptonic Flavor and CP Violations,''
Phys. Lett. B \textbf{723}, 107-112 (2013)
%doi:10.1016/j.physletb.2013.04.049
[arXiv:1303.0668 [hep-ph]].
%61 citations counted in INSPIRE as of 06 Feb 2021

%\cite{McKeen:2013dma}
\bibitem{McKeen:2013dma}
D.~McKeen, M.~Pospelov and A.~Ritz,
``Electric dipole moment signatures of PeV-scale superpartners,''
Phys. Rev. D \textbf{87}, no.11, 113002 (2013)
%doi:10.1103/PhysRevD.87.113002
[arXiv:1303.1172 [hep-ph]].
%77 citations counted in INSPIRE as of 06 Feb 2021

%\cite{Moroi:2013vya}
\bibitem{Moroi:2013vya}
T.~Moroi, M.~Nagai and T.~T.~Yanagida,
``Lepton Flavor Violations in High-Scale SUSY with Right-Handed Neutrinos,''
Phys. Lett. B \textbf{728}, 342-346 (2014)
%doi:10.1016/j.physletb.2013.11.058
[arXiv:1305.7357 [hep-ph]].
%18 citations counted in INSPIRE as of 06 Feb 2021

%\cite{Altmannshofer:2013lfa}
\bibitem{Altmannshofer:2013lfa}
W.~Altmannshofer, R.~Harnik and J.~Zupan,
``Low Energy Probes of PeV Scale Sfermions,''
JHEP \textbf{11}, 202 (2013)
%doi:10.1007/JHEP11(2013)202
[arXiv:1308.3653 [hep-ph]].
%78 citations counted in INSPIRE as of 06 Feb 2021

%\cite{Smith:2017dtz}
\bibitem{Smith:2017dtz}
C.~Smith and S.~Touati,
``Electric dipole moments with and beyond flavor invariants,''
Nucl. Phys. B \textbf{924}, 417-452 (2017)
%doi:10.1016/j.nuclphysb.2017.09.013
[arXiv:1707.06805 [hep-ph]].
%9 citations counted in INSPIRE as of 08 Feb 2021

%\cite{Evans:2018ewb}
\bibitem{Evans:2018ewb}
J.~L.~Evans, K.~Kadota and T.~Kuwahara,
``Revisiting Flavor and CP Violation in Supersymmetric $SU(5)$ with Right-Handed Neutrinos,''
Phys. Rev. D \textbf{98}, no.7, 075030 (2018)
%doi:10.1103/PhysRevD.98.075030
[arXiv:1807.08234 [hep-ph]].
%4 citations counted in INSPIRE as of 06 Feb 2021
 
%\cite{Maki:1962mu}
\bibitem{Maki:1962mu}
Z.~Maki, M.~Nakagawa and S.~Sakata,
``Remarks on the unified model of elementary particles,''
Prog. Theor. Phys. \textbf{28}, 870-880 (1962)
%doi:10.1143/PTP.28.870
%4249 citations counted in INSPIRE as of 06 Feb 2021

%\cite{Pontecorvo:1967fh}
\bibitem{Pontecorvo:1967fh}
B.~Pontecorvo,
``Neutrino Experiments and the Problem of Conservation of Leptonic Charge,''
Sov. Phys. JETP \textbf{26}, 984-988 (1968)
%2238 citations counted in INSPIRE as of 06 Feb 2021


%\cite{Buttazzo:2013uya}
\bibitem{Buttazzo:2013uya}
D.~Buttazzo, G.~Degrassi, P.~P.~Giardino, G.~F.~Giudice, F.~Sala, A.~Salvio and A.~Strumia,
``Investigating the near-criticality of the Higgs boson,''
JHEP \textbf{12}, 089 (2013)
%doi:10.1007/JHEP12(2013)089
[arXiv:1307.3536 [hep-ph]].
%1106 citations counted in INSPIRE as of 06 Feb 2021

%\cite{Bagnaschi:2014rsa}
\bibitem{Bagnaschi:2014rsa}
E.~Bagnaschi, G.~F.~Giudice, P.~Slavich and A.~Strumia,
``Higgs Mass and Unnatural Supersymmetry,''
JHEP \textbf{09}, 092 (2014)
%doi:10.1007/JHEP09(2014)092
[arXiv:1407.4081 [hep-ph]].
%169 citations counted in INSPIRE as of 06 Feb 2021
 
%\cite{Allanach:2001kg}
\bibitem{Allanach:2001kg}
B.~C.~Allanach,
``SOFTSUSY: a program for calculating supersymmetric spectra,''
Comput. Phys. Commun. \textbf{143}, 305-331 (2002)
%doi:10.1016/S0010-4655(01)00460-X
[arXiv:hep-ph/0104145 [hep-ph]].
%1236 citations counted in INSPIRE as of 06 Feb 2021
 
%\cite{Andreev:2018ayy}
\bibitem{Andreev:2018ayy}
V.~Andreev \textit{et al.} [ACME],
``Improved limit on the electric dipole moment of the electron,''
Nature \textbf{562}, no.7727, 355-360 (2018)
%doi:10.1038/s41586-018-0599-8
%278 citations counted in INSPIRE as of 06 Feb 2021
 
%\cite{TheMEG:2016wtm}
\bibitem{TheMEG:2016wtm}
A.~M.~Baldini \textit{et al.} [MEG],
``Search for the lepton flavour violating decay $\mu ^+ \rightarrow \mathrm {e}^+ \gamma $ with the full dataset of the MEG experiment,''
Eur. Phys. J. C \textbf{76}, no.8, 434 (2016)
%doi:10.1140/epjc/s10052-016-4271-x
[arXiv:1605.05081 [hep-ex]].
%573 citations counted in INSPIRE as of 06 Feb 2021

%\cite{Giudice:2003jh}
\bibitem{Giudice:2003jh}
G.~F.~Giudice, A.~Notari, M.~Raidal, A.~Riotto and A.~Strumia,
``Towards a complete theory of thermal leptogenesis in the SM and MSSM,''
Nucl. Phys. B \textbf{685}, 89-149 (2004)
%doi:10.1016/j.nuclphysb.2004.02.019
[arXiv:hep-ph/0310123 [hep-ph]].
%770 citations counted in INSPIRE as of 06 Feb 2021

%\cite{Buchmuller:2004nz}
\bibitem{Buchmuller:2004nz}
W.~Buchmuller, P.~Di Bari and M.~Plumacher,
``Leptogenesis for pedestrians,''
Annals Phys. \textbf{315}, 305-351 (2005)
%doi:10.1016/j.aop.2004.02.003
[arXiv:hep-ph/0401240 [hep-ph]].
%794 citations counted in INSPIRE as of 06 Feb 2021

\bibitem{HiraoMoroiFuture}
K.~Hirao and T.~Moroi,
work in progress.


\end{thebibliography}
\end{document}